\documentclass[11pt,a4paper]{article}
\usepackage{jheppub}
\usepackage{longtable}
\usepackage{bm}

\newcommand*{\no}{\noindent}
\newcommand*{\bea}{\begin{eqnarray}}
\newcommand*{\eea}{\end{eqnarray}}
\newcommand*{\be}{\begin{equation}}
\newcommand*{\ee}{\end{equation}}

\newcommand*{\pref}[1]{(\ref{#1})}

\newcommand*{\mn}{{\mu\nu}}

\newcommand*{\nn}{\nonumber}

\title{On the gauge boson's properties in a candidate technicolor theory}

\author{Axel Maas}
\affiliation{Institute for Theoretical Physics, Friedrich-Schiller-University Jena, Max-Wien-Platz 1, D-07743 Jena, Germany}
\emailAdd{axelmaas@web.de}

\abstract{The technicolor scenario replaces the Higgs sector of the standard model with a strongly interacting sector. One candidate for a realization of such a sector is two-technicolor Yang-Mills theory coupled to two degenerate flavors of adjoint, massless techniquarks.

Using lattice gauge theory the properties of the technigluons in this scenario are investigated as a function of the techniquark mass towards the massless limit. For that purpose the minimal Landau gauge two-point and three-point correlation functions are determined, including a detailed systematic error analysis.

The results are, within the relatively large systematic uncertainties, compatible with a behavior very similar to QCD at finite techniquark mass. However, the limit of massless techniquarks exhibits features which could be compatible with a (quasi-)conformal behavior.}


\begin{document}

\maketitle

\section{Introduction}

The standard-model Higgs particle is a very enigmatic object. Though being as simple a particle as possible, a fundamental scalar, it is unclear how it can stabilize its properties against quantum corrections \cite{Morrissey:2009tf}. Furthermore, the Higgs sector could not be viable at all due to triviality, except as a low-energy effective theory having an explicit cutoff \cite{Callaway:1988ya}. However, the description of standard-model processes with a Higgs degree of freedom is highly successful \cite{Bohm:2001yx}. Therefore, an effective Higgs degree of freedom as a low-energy approximation seems to be an important element of phenomenology.

This does not state anything about the high-energy properties of this effective low-energy degree of freedom \cite{Morrissey:2009tf}. As a consequence, many models have been developed which provide for such a degree of freedom, but without its particular problems. Among the theoretically most successful proposals are those which protect its properties with some custodial symmetry, like supersymmetry, make it an effective degree of freedom of a higher-dimensional underlying space-time manifold, or which make it a composite object \cite{Morrissey:2009tf}.

Among the latter proposals technicolor \cite{Sannino:2009za,Hill:2002ap,Lane:2002wv} has a number of intriguing features. In its simplest form, it replaces the Higgs by a bound state of two techniquarks, bound together by technigluons. The formal representation is that of a Yang-Mills theory coupled to a number of, at tree-level, massless techniquarks in some representation of the gauge group.

A requirement for technicolor is to be able to provide not only the masses for the weak gauge bosons but also for the fermions. This in turn requires technicolor to be walking and thus to be strongly interacting over a large range of energies, typical the electroweak scale up to some hundreds or possibly thousands of TeV \cite{Sannino:2009za,Hill:2002ap,Lane:2002wv}. Otherwise, there is no possibility to provide the vastly different fermion masses in the standard model, without the introduction of many new parameters or many additional degrees of freedom.

A possibility to achieve such a large domain of strong interactions is to have at those momenta a quasi-conformal behavior, i.\ e., the effective coupling is strong and almost energy-independent over the relevant domain. Possible candidates for theories with such a behavior have been identified and investigated with various methods \cite{Sannino:2009za}. A particular simple candidate for such a walking technicolor theory has a gauge-algebra su(2) with two flavors of adjoint techniquarks\footnote{Though it is neither clear whether technicolor exists nor whether the theory investigated here is indeed a possible realization of a technicolor scenario, the terms of technicolor, like techniquarks, will be retained for the quantities for the sake of simplicity and to avoid a proliferation of would-be.}. This theory has been investigated repeatedly \cite{Sannino:2009za,DelDebbio:2010hu,DelDebbio:2010hx,Bursa:2009we,DelDebbio:2009fd,Catterall:2008qk,Catterall:2007yx,Hietanen:2008mr,Hietanen:2009az,DeGrand:2009mt,DeGrand:2011qd,Lucini:2009an,Catterall:2009sb}. Though the theory shows strong interactions over a wide range of momenta, the results point more to a conformal theory rather than just a quasi-conformal theory. If this turns out to be the case, this conformality would have to be broken in order for the theory to be useful as an extension of the standard model. Whether this can be achieved by coupling it to the standard model, or if further sectors would be necessary, is yet not fully settled, and it may be that this is not possible at all. Nonetheless, at any finite techniquark mass, the theory cannot be genuinely conformal, but may still show a quasi-conformal behavior.

The aim here is to investigate the technigluon sector of this candidate theory to determine if and how such a (quasi-)conformal behavior manifests itself. To this end the gauge-dependent two-point and three-point correlation functions and the running gauge coupling will be determined using lattice gauge theory. Since in lattice gauge theory it is not yet possible to treat massless fermions numerically, massive techniquarks will be used instead, and their limit to zero mass will be studied.

The use of the gauge-dependent correlation functions is complementary to the existing calculations of gauge-invariant quantities. The advantage is the possibility to access the energy dependence of both particle properties and interactions. This permits a very direct test of the distinct behavior such quantities would exhibit in a (quasi-)conformal theory, in particular the running gauge coupling \cite{Sannino:2009za,Frishman:2010zz}. Furthermore, these quantities are valuable input for functional calculations of gauge-invariant quantities, which can be performed at zero techniquark mass. Such a strategy has been successfully applied in QCD \cite{Alkofer:2000wg,Fischer:2006ub,Maas:2010gi,Alkofer:2008et,Braun:2009gm,Fischer:2010fx,Blank:2010pa,Pawlowski:2010ht,Binosi:2009qm,Dudal:2010cd}, and is also a promising approach to technicolor \cite{Elias:1984zh,Gies:2005as,Braun:2009ns,Aguilar:2010ad,Sannino:2009za,Doff:2009na,Braun:2010qs}.

Details of the setup of the investigation are provided in section \ref{sec:setup}. The present investigation is based on configurations provided by the authors of the papers \cite{DelDebbio:2010hu,DelDebbio:2010hx,DelDebbio:2009fd,DelDebbio:2008zf}, where also more details on the generation and the properties of the configurations can be found. It is of further relevance how the scale is set for the calculations here. One possibility will be discussed in detail in section \ref{sec:scale}. There are many other sources of systematic uncertainties in such calculations. These will be discussed at length in section \ref{sec:sys}. The final results will be given in section \ref{sec:res}. A discussion and a possible interpretation of the results will be given in section \ref{sec:inter}. In addition, a comparison to Yang-Mills theory will be made throughout to assess the influence of the techniquarks. Some concluding remarks will be made in section \ref{sec:conc}.

\section{Setup}\label{sec:setup}

The basic setup for the present lattice calculation is a Wilson action for the gauge sector coupled to two degenerate Wilson fermions in the adjoint representation. The configurations have been generated with the methods described in \cite{DelDebbio:2008zf} and produced and provided by the authors of this paper. These configurations have already been employed in various calculations of the spectrum and other gauge-invariant quantities, see \cite{DelDebbio:2010hu,DelDebbio:2010hx,DelDebbio:2009fd}. The latter results will also be used in section \ref{sec:scale}.

\begin{longtable}{|c|c|c|c|c|c|}
\caption{\label{beta}The configurations employed. $N_t$ is the temporal and $N_s$ the spatial extent of the lattice, and the total volume in lattice units is $N_t\times N_s^3$. The bare techniquark mass $am_0$ influences the techniquark mass, and is absent in Yang-Mills theory. For the setting of the scale, see section \ref{sec:scale}. If two numbers are given for configurations then the first number denotes the number of configurations used for the determination of the gauge dependency and the second number the one for the final results. Otherwise, the same number of configurations has been used.}\\
 \hline
 $-am_0$ & $\beta$ & $a^{-1}$ [TeV] & $N_t$ & $N_s$ & Configurations \endfirsthead
 \hline
 \multicolumn{6}{|l|}{Table \ref{beta} continued}\\
 \hline
 $-am_0$ & $\beta$ & $a^{-1}$ [TeV] & $N_t$ & $N_s$ & Configurations \endhead
 \hline
 \multicolumn{6}{|r|}{Continued on next page}\\
 \hline\endfoot
 \endlastfoot
 \hline
 \hline
 -0.5 & 2.25 & 1.66 & 16 & 8 & 721 \cr
 \hline
 -0.25 & 2.25 & 1.73 & 16 & 8 & 720 \cr
 \hline
 0 & 2.25 & 1.76 & 16 & 8 & 720 \cr
 \hline
 0.25 & 2.25 & 1.81 & 16 & 8 & 721 \cr
 \hline
 0.5 & 2.25 & 1.91 & 16 & 8 & 721 \cr
 \hline
 0.75 & 2.25 & 2.18 & 16 & 8 & 721 \cr
 \hline
 0.90 & 2.25 & 2.22 & 16 & 8 & 721 \cr
 \hline
 0.95 & 2.25 & 2.52 & 16 & 8 & 721 \cr
 \hline
 0.95 & 2.25 & 2.52 & 24 & 12 & 329 \cr
 \hline
 0.975 & 2.25 & 2.74 & 16 & 8 & 721 \cr
 \hline
 1.00 & 2.25 & 3.04 & 16 & 8 & 721 \cr
 \hline
 1.00 & 2.25 & 3.04 & 24 & 12 & 280 \cr
 \hline
 1.025 & 2.25 & 3.42 & 16 & 8 & 721 \cr
 \hline
 1.05 & 2.25 & 3.92 & 16 & 8 & 721 \cr
 \hline
 1.05 & 2.25 & 3.92 & 24 & 12 & 54 \cr
 \hline
 1.075 & 2.25 & 4.59 & 16 & 8 & 180 \cr
 \hline
 1.075 & 2.25 & 4.59 & 24 & 12 & 53 \cr
 \hline
 1.1 & 2.25 & 5.51 & 16 & 8 & 180 \cr
 \hline
 1.1 & 2.25 & 5.51 & 24 & 12 & 54 \cr
 \hline
 1.125 & 2.25 & 6.79 & 16 & 8 & 195 \cr
 \hline
 1.125 & 2.25 & 6.79 & 24 & 12 & 55 \cr
 \hline
 1.15 & 2.25 & 8.63 & 16 & 8 & 150 \cr
 \hline
 1.15 & 2.25 & 8.63 & 32 & 16 & 96 \cr
 \hline
 1.175 & 2.25 & 11.3 & 16 & 8 & 150 \cr
 \hline
 1.175 & 2.25 & 11.3 & 24 & 12 & 161 \cr
 \hline
 1.175 & 2.25 & 11.3 & 24 & 24 & 39 \cr
 \hline
 1.175 & 2.25 & 11.3 & 32 & 16 & 92 \cr
 \hline
 1.18 & 2.25 & 12.0 & 24 & 12 & 172 \cr
 \hline
 1.18 & 2.25 & 12.0 & 32 & 16 & 141 \cr
 \hline
 1.18 & 2.25 & 12.0 & 64 & 24 & 13 \cr
 \hline
 1.185 & 2.25 & 12.8 & 24 & 12 & 167\cr
 \hline
 1.185 & 2.25 & 12.8 & 32 & 16 & 15\cr
 \hline
 1.185 & 2.25 & 12.8 & 64 & 24 & 12 \cr
 \hline
 1.19 & 2.25 & 13.6 & 24 & 12 & 167 \cr
 \hline
 1.19 & 2.25 & 13.6 & 32 & 16 & 45 \cr
 \hline
 1.19 & 2.25 & 13.6 & 64 & 24 & 12 \cr
 \hline
 1.2 & 2.25 & 15.4 & 16 & 8 & 180 \cr
 \hline
 1.2 & 2.25 & 15.4 & 24 & 12 & 167 \cr
 \hline
 1.2 & 2.25 & 15.4 & 32 & 16 & 45 \cr
 \hline
 \hline
  & 2.25 & 1.48 & 16 & 8 & 861\cr
 \hline
  & 2.25 & 1.48 & 24 & 12 & 37/130\cr
 \hline
  & 2.25 & 1.48 & 24 & 24 & 145/145 \cr
 \hline
  & 2.25 & 1.48 & 32 & 16 & 125/281 \cr
 \hline
  & 2.25 & 1.48 & 64 & 16 & 87/58 \cr
 \hline
  & 2.25 & 1.48 & 64 & 24 & 27/51 \cr
 \hline
  & 2.538 & 3.62 & 16 & 8 & 222\cr
 \hline
  & 2.636 & 5.12 & 24 & 12 & 145\cr
 \hline
  & 2.636 & 5.12 & 24 & 24 & 37/145 \cr
 \hline
  & 2.736 & 7.15 & 32 & 16 & 125/281 \cr
 \hline
  & 2.736 & 7.15 & 64 & 16 & 96/135 \cr
 \hline
  & 2.736 & 7.15 & 64 & 24 & 34/12 \cr
 \hline
\end{longtable}

As noted in the introduction, it is not (yet) possible to simulate such a theory with massless fermions. Therefore, a set of configurations with differing techniquark masses, implemented by different hopping parameters $\kappa$, has been studied. For the present matter content and a $\beta$ of 2.25, the critical $\kappa$ may be such that the bare techniquark mass is $am_0\approx -1.2$ \cite{DelDebbio:2010hu,DelDebbio:2010hx,DelDebbio:2009fd}. The different setups included are listed in table \ref{beta}. Note that only a subset of the configurations employed in calculations of the spectrum could be used, as the quantities studied here are suspected to have a longer auto-correlation time than the masses of low-lying gauge-invariant excitations \cite{Maas:2008ri}. For comparison also configurations of pure Yang-Mills theory have been generated, using the same Wilson action. Details of their generation can be found in \cite{Cucchieri:2006tf}.

The study of the technigluon, and thus the gauge boson, requires to fix the gauge. The gauge chosen here is Landau gauge, which is comparatively well understood in the case of Yang-Mills theory. It has furthermore a number of technical advantages \cite{Fischer:2008uz}, in particular very simple renormalization properties. However, in the strongly-interacting regime desired for a technicolor theory, gauge fixing becomes more complicated due to the Gribov-Singer ambiguity \cite{Gribov:1977wm,Singer:1978dk}. Thus, the term Landau gauge does in general not describe a unique gauge choice, and has to be specified further. Here, this ambiguity will be lifted by choosing the minimal Landau gauge \cite{Cucchieri:1995pn}. Some alternative choices will be discussed in section \ref{sec:gauge}. It turns out that within the systematic uncertainties of the present investigations this choice is not of relevance.

The definition of minimal Landau gauge, as that of Landau gauge itself, only involves the gauge boson fields. Fixing to Landau gauge can thus be done in the same way as in Yang-Mills theory. The method employed here is described in \cite{Cucchieri:2006tf}. The presence of the matter fields turns out not to have any unexpected effects on the numerical behavior of the gauge fixing.

With the gauge-fixed configurations available it remains to calculate the correlation functions. Here only the gauge sector will be studied, up to and including three-point functions. Since Lorentz symmetry is manifest in Landau gauge all one-point functions vanish. The simplest non-trivial ones are therefore two-point functions.

One is the technigluon propagator, $D^{ab}_\mn$, which takes the form
\be
D^{ab}_\mn=\delta^{ab}\left(\delta_\mn-\frac{p_\mu p_\nu}{p^2}\right)D(p)\label{gluonprop},
\ee
\no with the dressing function $Z(p)=p^2D(p)$. The propagator is assumed to be color-diagonal, which at least for Yang-Mills theory is well fulfilled \cite{Maas:2010qw}. In Landau gauge auxiliary degrees of freedom are introduced, the Faddeev-Popov technighost fields. Their propagator is given by
\be
D^{ab}_G=\delta^{ab}D_G(p)=-\delta^{ab}\frac{G(p)}{p^2}\nn.
\ee
\no Its color diagonality is an exact consequence of its equation of motion \cite{Maas:2010qw}. Both propagators are determined using the methods described in \cite{Cucchieri:2006tf}.

An important property of Landau gauge is that the running gauge coupling in the so-called miniMOM scheme \cite{vonSmekal:2009ae}, which can be translated into the $\overline{\mathrm{MS}}$ scheme, can be determined just from these two renormalized propagators as \cite{vonSmekal:1997vx}
\be
\alpha(p^2)=\alpha(\mu^2)G(p,\mu)^2Z(p,\mu)\label{alpha}.
\ee
\no In the perturbative domain, this coincides with the usual running coupling.

There are two three-point vertices connecting the technigluon and technighost fields. One is the technighost-technigluon vertex $\Gamma^{abc}_\mu$ and the other the three-technigluon vertex $\Gamma^{abc}_{\mu\nu\rho}$. In a lattice computation only the non-amputated vertices can be determined \cite{Cucchieri:2006tf}. Thus, in Landau gauge only the tensor components transverse in the technigluon momenta can be obtained, but these are actually the only relevant ones \cite{Fischer:2008uz}. This leaves still two color tensors for the technighost-technigluon vertex. However, for su(2) the second one vanishes identically since the symmetric color tensor is identically zero \cite{Cvitanovic:2008}. This leaves one transverse tensor component $A(p,q,p+q)$. It is a function of the technighost and the technigluon momentum, having an exchange symmetry on the technighost-antitechnighost legs \cite{Alkofer:2000wg}. It can be determined using the methods described in \cite{Cucchieri:2006tf,Cucchieri:2004sq}. Of the eight remaining color and Lorentz tensors of the three-technigluon vertex, four vanish again for su(2), leaving four \cite{Ball:1980ax}. Using the methods described in \cite{Cucchieri:2006tf}, any tensor component could be determined. Here, for reasons of limited statistics, only the component along the tree-level vertex will be determined, $B(p,q,p+q)$. It is totally Bose-symmetric \cite{Ball:1980ax}, and can thus be taken as a function of two arbitrary technigluon momenta.

Because the lattices available are mostly asymmetric, only one momentum configuration for the vertices will be investigated. This is the one where one of the technigluon momenta vanishes, i.\ e., in case of the technighost-technigluon vertex the only one and an arbitrary one for the three-technigluon vertex. The remaining momenta are then oriented along the extended lattice side. This permits to span the largest interval of momenta.

It thus remains to renormalize these four correlation functions, while the running coupling is a renormalization-group invariant. In Landau gauge, the dependency on the cutoff is at most logarithmic, and thus the expected effects are small, as is confirmed by the calculations below. Furthermore, it turns out that it is very complicated to determine a common renormalization scale for all systems due to the vastly different values of $a$ encountered. The precise renormalization prescription employed will be detailed in every instance.

Because of the relation \pref{alpha} the renormalization of both propagators is related \cite{vonSmekal:1997vx}. Requiring thus $G(\mu,\mu)=1$ fixes immediately also the renormalization of the the technigluon propagator. Concerning the vertices, both renormalization constants are finite \cite{Alkofer:2000wg}. The momentum configuration used for the renormalization purpose will be the asymmetric point $p^2=\mu^2$ and $q^2=0$. Using this point instead of the symmetric point is useful when comparing results to lower-dimensional lattice systems where the hypercubic symmetry makes it complicated to access a symmetric momentum point \cite{Maas:2007uv}. This completes the setup of the calculations.

\section{Scale setting}\label{sec:scale}

There are three significant complications in any attempt to set the scale for the present investigation.

One is that the only possible quantity which could possibly be assigned an at least speculative value from experiment is the mass of the would-be Higgs boson, a techniquark bound state \cite{Sannino:2009za,Hill:2002ap,Lane:2002wv}. However, its mass becomes infinite when making the techniquarks heavier and heavier, and a comparison to the quenched case with this scale is not meaningful. Thus, in the following a quantity will be used which has a finite value for Yang-Mills theory and for all systems investigated here. To avoid as many complications as possible, this will be the $0^{++}$ techniglueball mass, which is arbitrarily set to 2 TeV.

The second complication is that any value of the techniquark mass corresponds, in principle, to a different theory. Thus a direct comparison between setups with different techniquarks is meaningful only in a very specific sense. In particular, the ratio of the scale-setting techniglueball mass to the one of any other bound state can be different for differing techniquark mass. Hence, the comparison for differing techniquark mass should be seen as an investigation of a theory sequence rather than a comparison between individual theories.

Finally, the most cumbersome problem for the present investigation is actually the basic question, whether the theory is showing a (quasi-)conformal behavior in the limit of vanishing techniquark mass. If the theory were only walking, i.\ e., showing only a quasi-conformal behavior over a certain momentum range, this is not a problem in principle, as a scale setting can then be achieved outside the conformal domain, in particular the deep infrared. If the theory should turn out to be genuinely conformal, however, it has to be scale-free in the limit of vanishing techniquark mass \cite{Frishman:2010zz}. In this case there is no meaningful way of setting a scale at all. With the present way of setting the scale this should manifest itself by the fact that all bound state masses should vanish with the techniquark mass, and in fact with a characteristic behavior. As a consequence, the scale set by the techniglueball mass would have to diverge with the techniquark mass going to zero to keep the scale fixed.

\begin{figure}
\includegraphics[width=\linewidth]{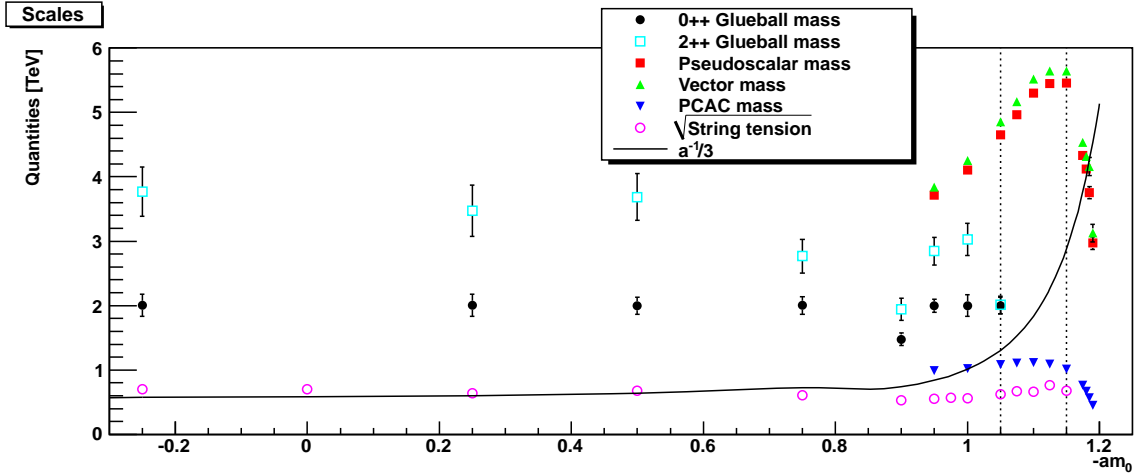}
\caption{\label{fig:scale}The scale $a^{-1}$ in TeV as a function of the hopping parameter $\kappa$. Alongside shown are the masses of lowest pseudo-scalar and scalar techniquark bound states, the PCAC techniquark mass \cite{DelDebbio:2010hu}, the masses of the $0^{++}$ and the $2^{++}$ glueballs and the root of the string tension \cite{DelDebbio:2010hx} in these units. The two vertical dotted lines denote the critical hopping parameter for a spurious phase transition on a $16\times 8^3$ and a $24\times 12^3$ lattice \cite{DelDebbio:2010hx}, see section \ref{sec:mass} for details.}
\end{figure}

Before discussing which of these potential problems play a role, it is useful to determine the actual scale. The result, using an interpolation/extrapolation\footnote{Note that a continuous polynomial has been used, and therefore one exceptional case at $am_0=-0.9$ did not yield precisely 2 TeV for the $0^{++}$ techniglueball after using the resulting formula. It appears that it is possible that this is a single statistical anomaly.} from the results published in \cite{DelDebbio:2010hx,DelDebbio:2010hu}, are shown in figure \ref{fig:scale}.

The so obtained scale shows a striking feature: It appears to diverge when the hopping parameter tends towards the critical value, though the $\beta$ value is fixed. This is suggestive of a second-order phase transition. This possibility will be discussed in detail after presenting the main results of this paper in section \ref{sec:inter}. Otherwise the technihadronic observables, including the PCAC mass of the techniquark, seem to vanish when approaching zero techniquark mass. Furthermore, essentially the same scale would have been obtained if the more conventional string tension would had been used instead, and in particular no vanishing of the techniglueball mass nor the string tension is visible. In particular, in these units the techniquark-mass-dependencies of the gluonic observables are much less pronounced than the hadronic ones, thus being an acceptable tool for performing this investigation in the sense of a limit of theories. Thus, it seems that the scale can be used for the intended purpose, to take the limit of theories. However, it still does not make sense to compare two systems with differing techniquark masses as independent theories, as they may have a different hierarchy of bound states masses.

It is an interesting aside that there must exist a bare techniquark mass larger than minus one for which the ratio of technihadron masses to the string tension must be minimal, before it rises again towards infinity when the techniquark mass is tuned to infinity and a Yang-Mills theory is obtained.

For completeness, the central results of this work will also be reported in lattice units, which essentially corresponds to setting the same scale for all systems. This also offers a different possible interpretation of the results, see section \ref{sec:inter}.

\section{Lattice artifacts}\label{sec:sys}

The lattice investigations performed here have a number of inherent limitations, which manifest themselves as systematic uncertainties. These will be investigated in turn. Of course, since there is no full systematic control over the artifacts, the following can only indicate a lower limit for the systematic uncertainties.

\subsection{Asymmetry effects}\label{sec:asy}

Most of the quantities investigated here have a Lorentz index, except for the ghost propagator. In case of asymmetric lattices, it is known that this can cause additional artifacts, in particular in the infrared \cite{Cucchieri:2006za,Cucchieri:2007ta}. In section \ref{sec:res} results will be shown as a function of momenta along the elongated axis. To estimate the influence of the asymmetry, a comparison will be made here for the same momentum configuration on a symmetric and an asymmetric lattice. This will be done using a fixed value of $am_0=-1.175$, for which asymmetric lattices of size $24\times 12^3$ and symmetric lattices of size $24^4$ are available. More extensive comparisons can be found for Yang-Mills theory, see \cite{Cucchieri:2006za,Silva:2005hb}. The latter results are qualitatively similar to the ones found here for the case with techniquarks.

\begin{figure}
\includegraphics[width=0.5\linewidth]{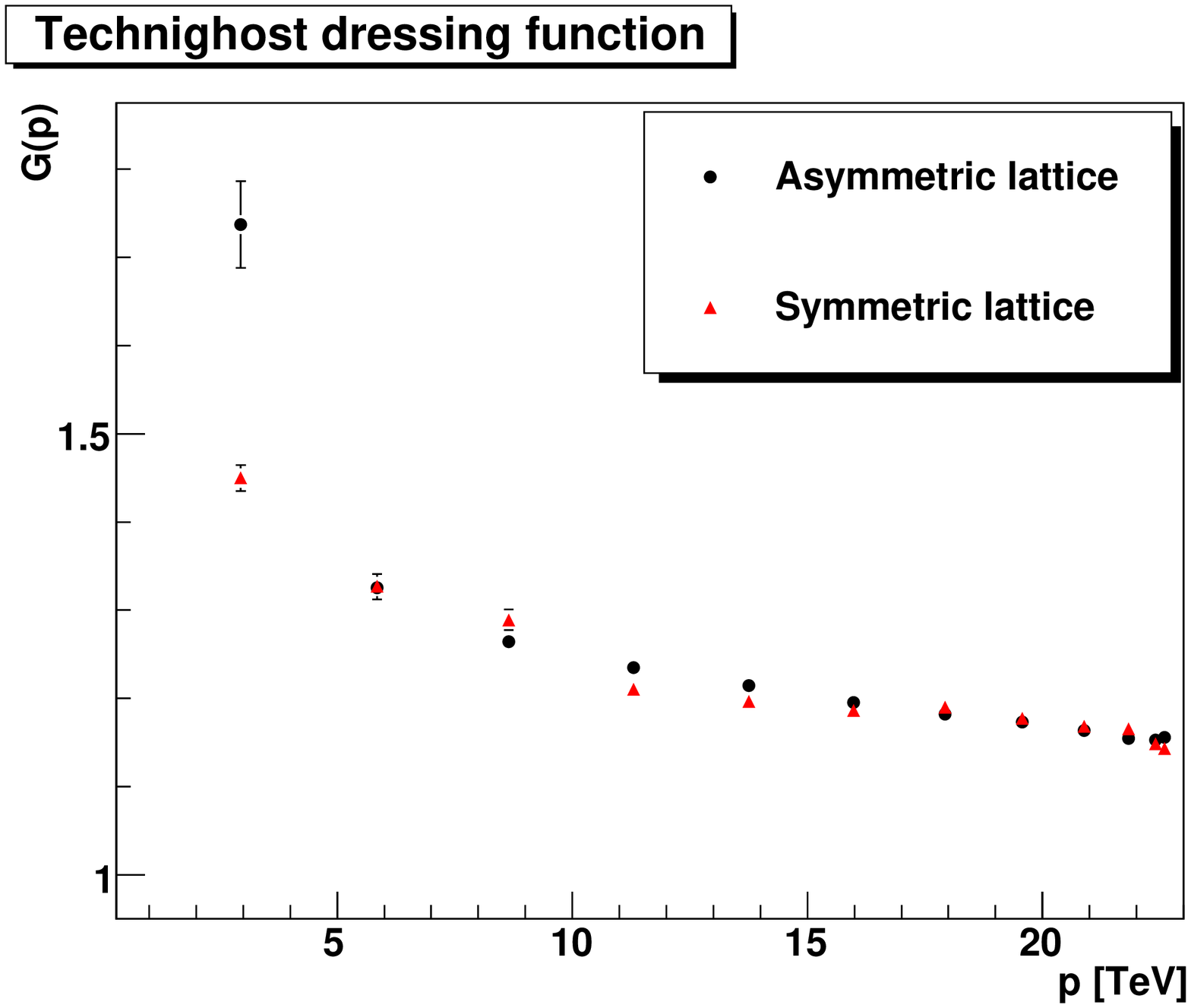}\includegraphics[width=0.5\linewidth]{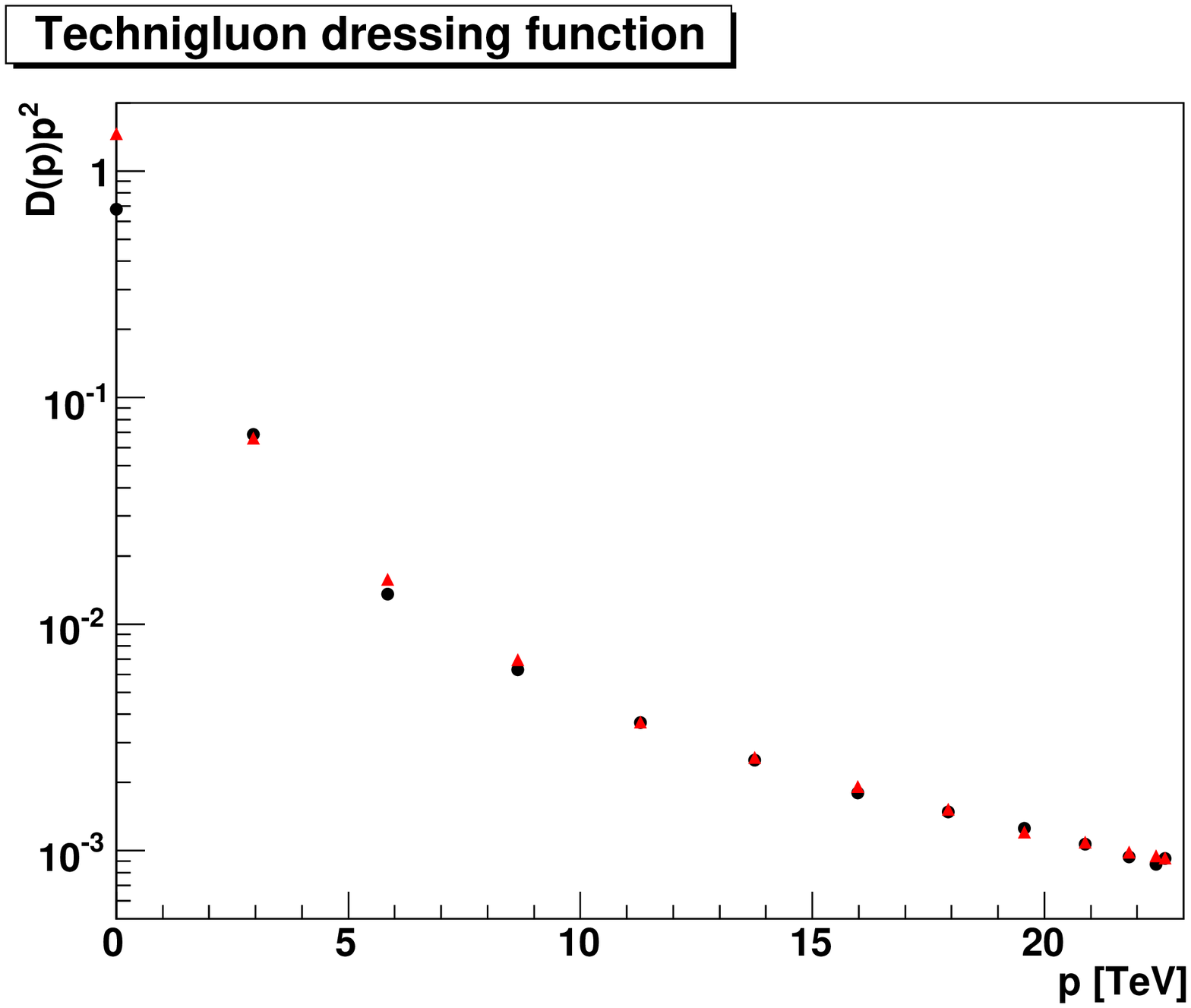}
\caption{\label{fig:prop-asy}The dressing functions compared for symmetric and asymmetric lattices with $am_0=-1.175$, and $24^4$ and $24\times 12^3$ lattice sites, respectively. The technighost is shown in the left panel and the technigluon in the right panel. The functions are not renormalized.}
\end{figure}

The simplest object is the technighost dressing function\footnote{Note that generically the technighost dressing function and the technigluon propagator have been found to be the quantities exhibiting the strongest sensitivity to lattice artifacts \cite{Cucchieri:2008fc,Cucchieri:2007rg,Bornyakov:2009ug,Bogolubsky:2009dc,Fischer:2007pf}. Therefore, these two quantities are always plotted, as it turns out to also apply to the current theory.}. It is plotted for the symmetric and asymmetric case in figure \ref{fig:prop-asy}. It is seen that for all but the lowest momentum point there is little difference between the symmetric and the asymmetric lattice. Since the propagators are rather sensitive to finite-volume corrections in the far infrared \cite{Cucchieri:2008fc,Cucchieri:2007rg,Bornyakov:2009ug,Bogolubsky:2009dc,Fischer:2007pf}, this may also be due to the different physical volumes. 

The situation for the technigluon is significantly more complicated. Since the technigluon is a vector particle, it can be polarized along the elongated edge or transverse to it \cite{Cucchieri:2007ta}. The behavior of the propagator for both polarization directions can be different. Both coincide by construction if the lattice is symmetric, and then also with the one defined in \pref{gluonprop}. Since the present investigation should be understood as an intermediate step towards the infinite volume case, the difference between the different polarization tensors is rather a finite volume effect than genuine. Therefore, here not the individual polarizations but their averaged behavior will be investigated. To assess again the systematic uncertainty emerging from this approximation, the difference between a symmetric and an asymmetric lattice is also shown in figure \ref{fig:prop-asy}. Within the (relatively large) statistical errors, no significant change when going from a symmetric lattice to an asymmetric lattice is visible, except at zero momentum. Therefore, the approximation appears to be acceptable for finite momenta. Again, finite-volume effects may distort this result.

Concerning the propagators, it seems therefore that any effect due to the asymmetry of the lattice is within the region where finite volume effects are expected to be strongest, but no additional severe effect is observed.

\begin{figure}
\includegraphics[width=\linewidth]{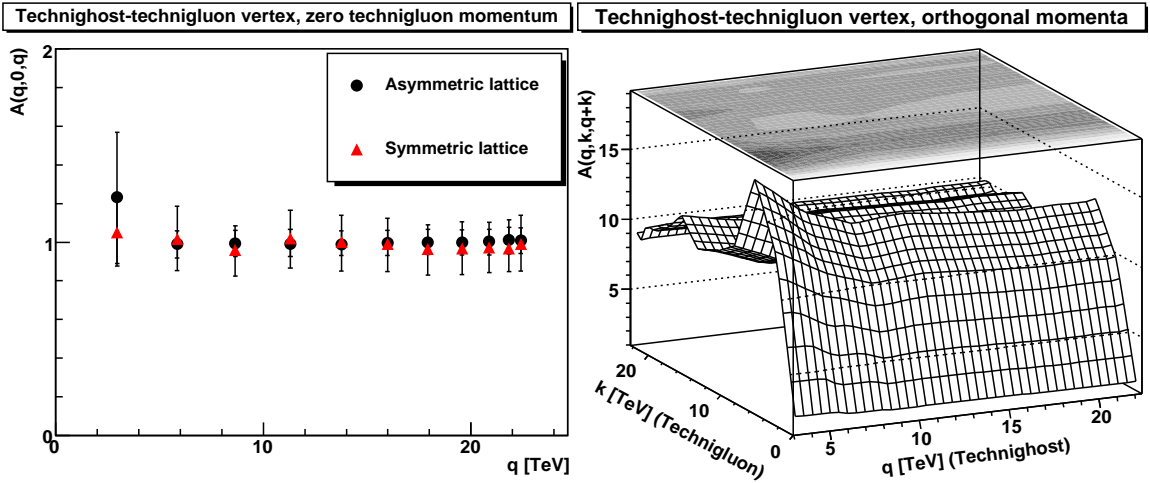}
\caption{\label{fig:vert-asy}The dressing function of the technighost-technigluon vertex compared for symmetric and asymmetric lattices with $am_0=-1.175$, and $24^4$ and $24\times 12^3$ lattice sites, respectively. In the left panel the momentum configuration $qk=0$ and $|k|=0$ is shown. The right panel shows, only for the symmetric case, the momentum configuration $qk=0$, with suppressed statistical errors.}
\end{figure}

It is worthwhile to also investigate the effects on the vertices, since they mix in general polarizations along different directions. Unfortunately, due to the excessive statistical noise \cite{Cucchieri:2006tf}, it was not possible to obtain conclusive results for the three-technigluon vertex. Here therefore only the technighost-technigluon-vertex will be investigated. Furthermore, while the three-technigluon vertex is Bose-symmetric, this is not the case for the technighost-technigluon vertex. It is therefore worthwhile to check in the symmetric case to which extent it changes when using off-linear momentum configurations. Both results are shown in figure \ref{fig:vert-asy}.

First of all, for the momentum configuration which will be used below there is, within statistical uncertainty, no difference between the symmetric and the asymmetric case. Thus an investigation of the asymmetric case is sensible. Much more dramatic is the situation for the other momentum configuration in the symmetric case. There, a strong suppression of the vertex for small technigluon momenta is seen. This is significantly different from Yang-Mills theory, where the vertex is essentially flat for all momenta \cite{Cucchieri:2008qm,Schleifenbaum:2004id,Ilgenfritz:2006he}. This can have dramatic influence on approximation schemes in functional calculations \cite{Alkofer:2000wg,Fischer:2009tn,Alkofer:2008jy}, even when investigating only the case at finite adjoint quark mass \cite{Aguilar:2010ad}, and may have very interesting consequences. This certainly deserves further attention.

\subsection{Volume dependence}\label{sec:volume}

Finite-volume artifacts probably have the strongest influence on the low-momentum properties of the gauge sector in Yang-Mills theory \cite{Cucchieri:2008fc,Cucchieri:2007rg,Bornyakov:2009ug,Bogolubsky:2009dc,Fischer:2007pf}. It can therefore be expected that such strong effects are also present when matter fields are included. It is therefore of significant importance to compare results for different volumes at fixed $a$ (and here thus fixed $\kappa$). This will be done for two different bare techniquark masses values, one will be $-1.175$ and one will be $-0.95$, to investigate the situation both  far away and close to the critical case. Note that in all cases the same aspect ratio is maintained, to evade possible cross-contamination effects.

\begin{figure}
\includegraphics[width=0.5\linewidth]{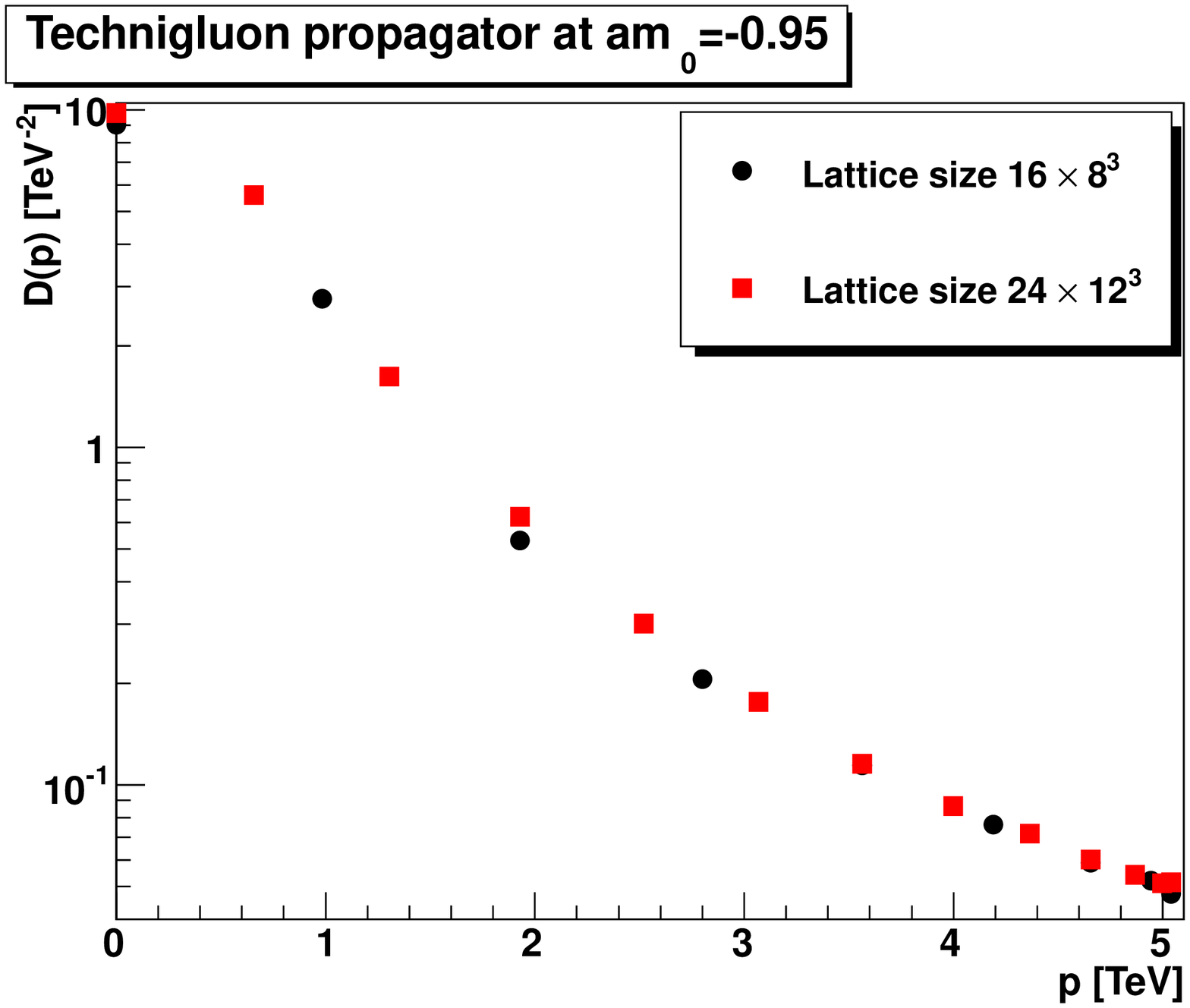}\includegraphics[width=0.5\linewidth]{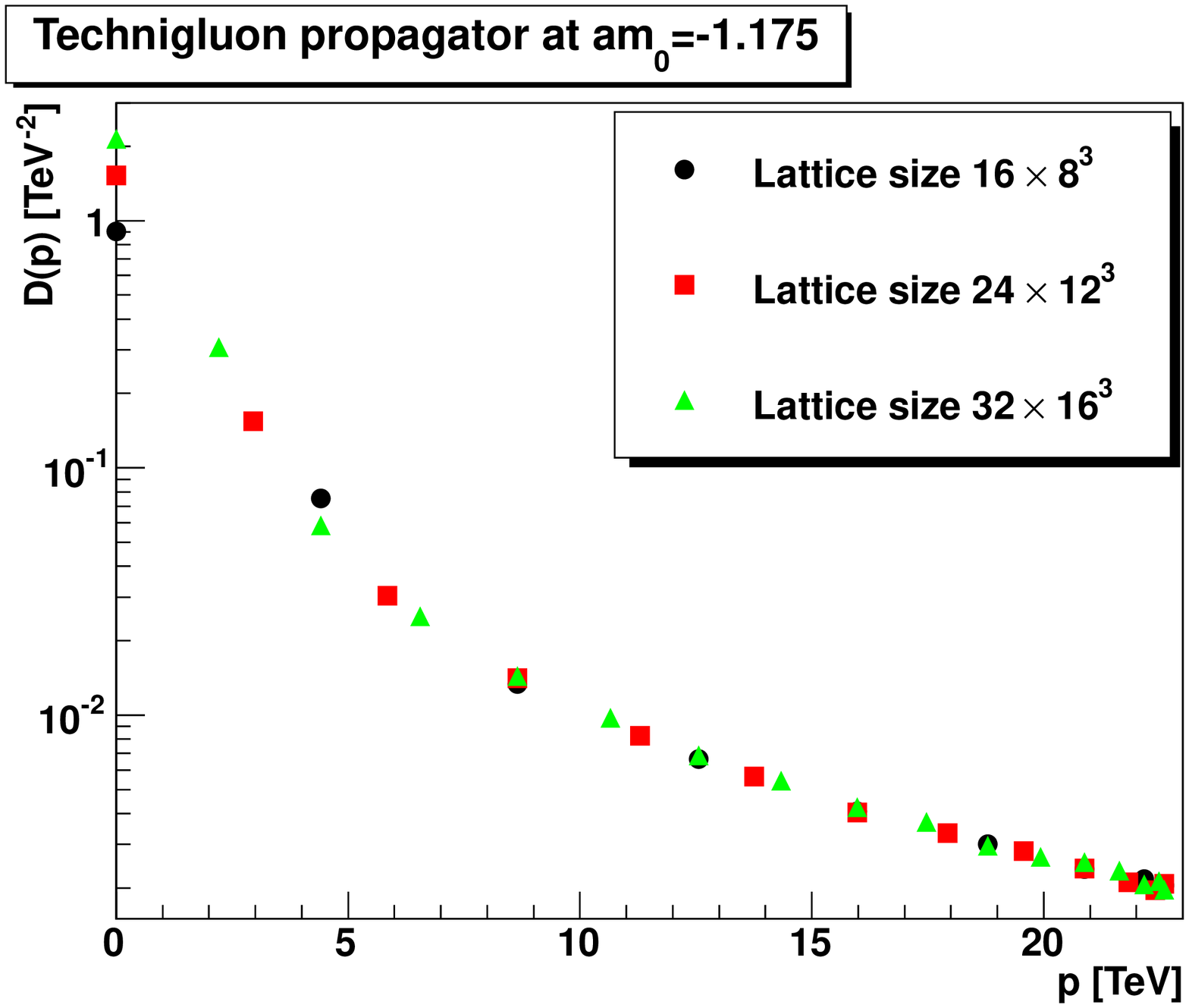}
\caption{\label{fig:gp-v}The technigluon propagator as a function of the lattice size at otherwise fixed parameters. The left-hand panel shows the results for $am_0=-0.95$, the right-hand panel shows the result for $am_0=-1.175$. Results are not renormalized.}
\end{figure}

The first quantity to check is the technigluon propagator. The results are shown in figure \ref{fig:gp-v} for the two selected bare techniquark masses. As expected, the most significant finite-volume artifacts are visible in the infrared. In particular, at zero momentum the change can be a factor of more than two, making this point unreliable\footnote{This volume-dependence is interesting in its own right \cite{Fischer:2007pf}, but would require many more different volumes than presently available to make use of.}. The lowest non-zero momentum point also still shows variations between the different volumes, though the effect is much smaller, at most of the order of 30\%. For even larger momenta, the finite-volume effects appear to be negligible in the present context.

\begin{figure}
\includegraphics[width=0.5\linewidth]{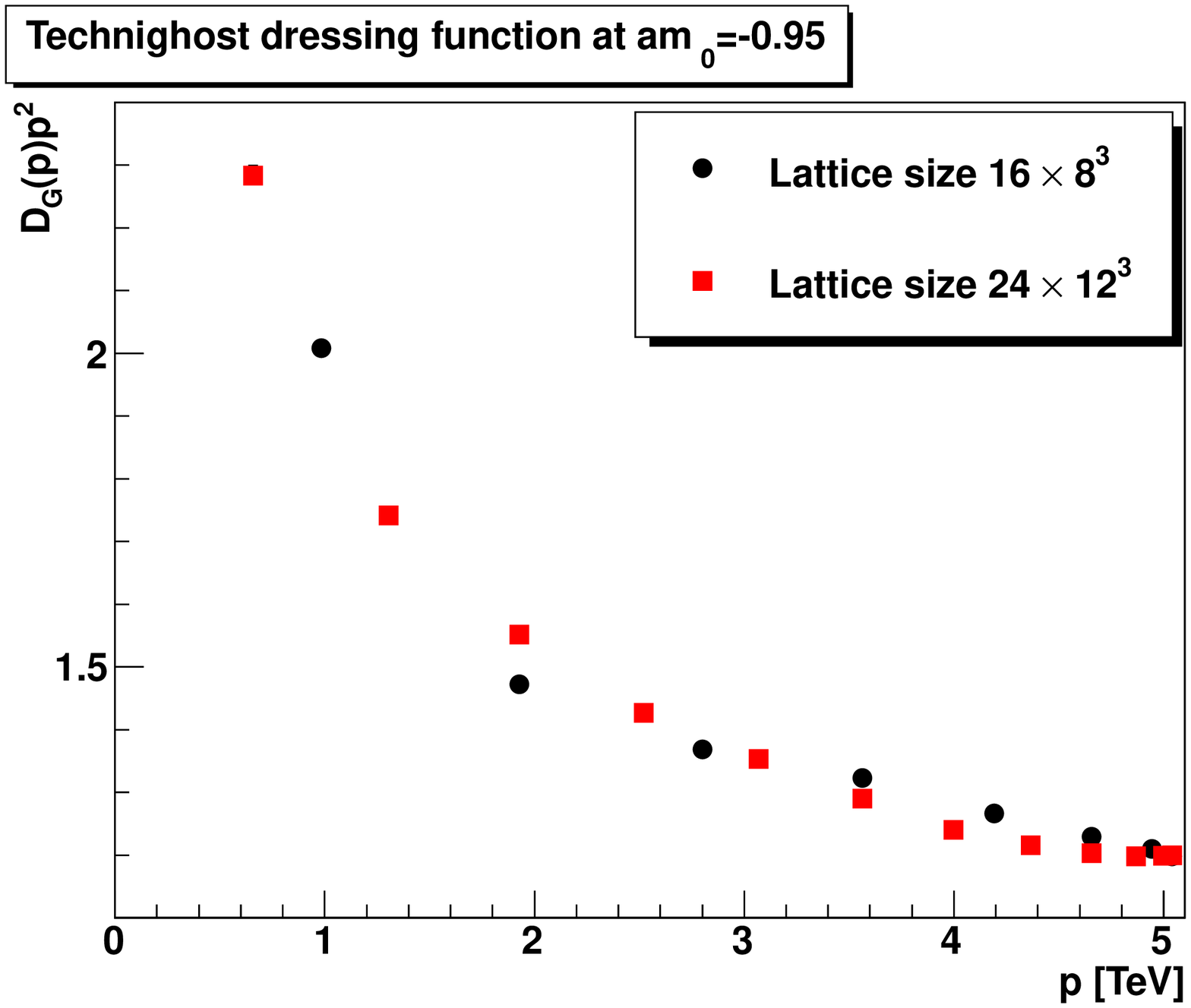}\includegraphics[width=0.5\linewidth]{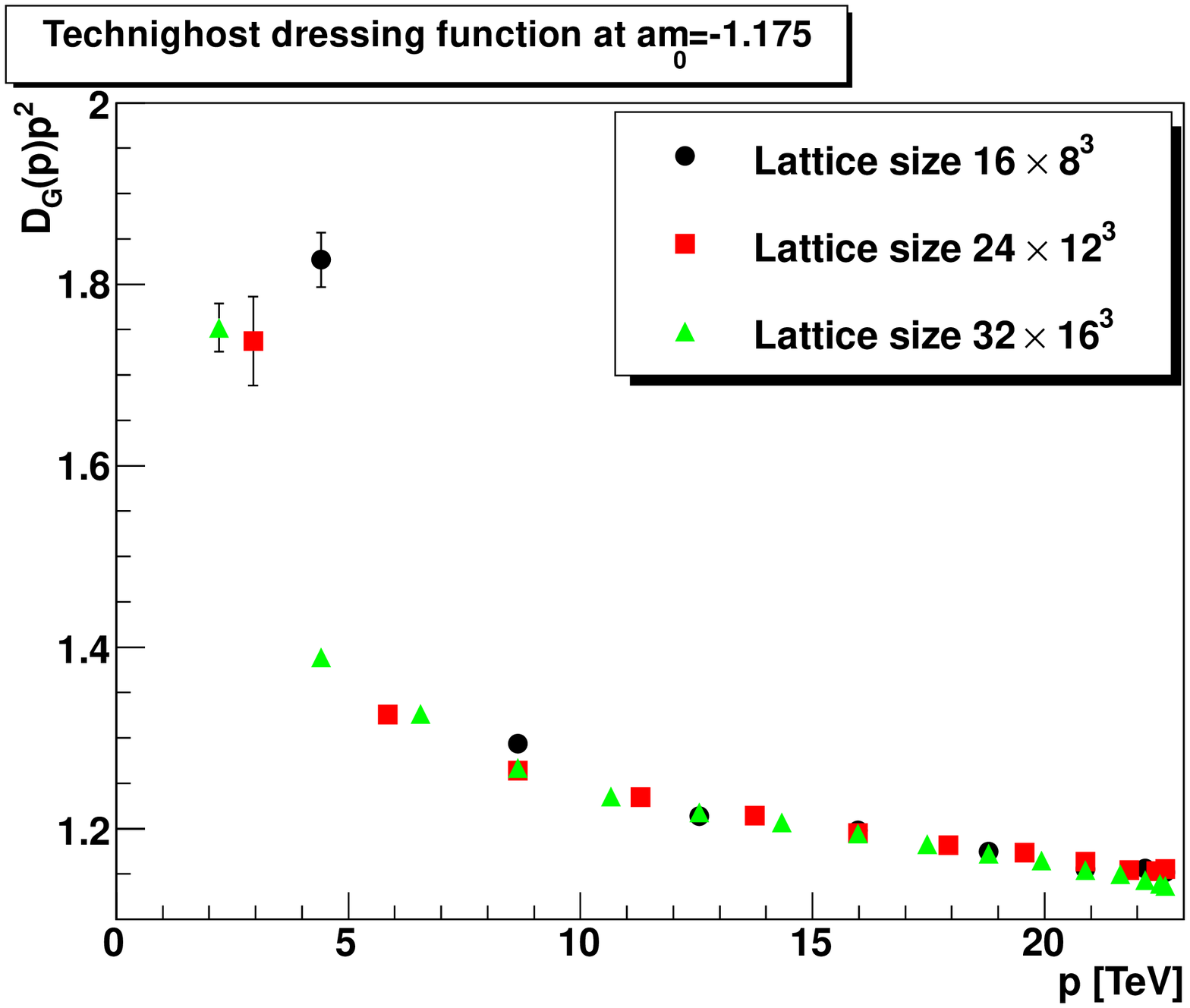}
\caption{\label{fig:ghp-v}The technighost dressing function as a function of the lattice size at otherwise fixed parameters. The left-hand panel shows the results for $am_0=-0.95$, the right-hand panel shows the result for $am_0=-1.175$. Results are not renormalized.}
\end{figure}

The second quantity is the technighost propagator, shown in figure \ref{fig:ghp-v}. The results indicate that the volume-dependence has two traits. The farther away from the critical point, the larger are the maximum momenta at which finite-volume effects are visible. The second is that the closer to the critical point the stronger the finite-volume-effects become in the infrared. Still, except for the lowest momentum closer to the critical point the finite-volume effects are only on the order of a few percent, and thus negligible for the present purpose. Only in case of the lowest momentum accessible\footnote{The ghost propagator cannot be determined for zero momentum on a finite lattice \cite{Cucchieri:1997ns}.}, significant finite-volume effects have to be taken into account.

\begin{figure}
\includegraphics[width=0.5\linewidth]{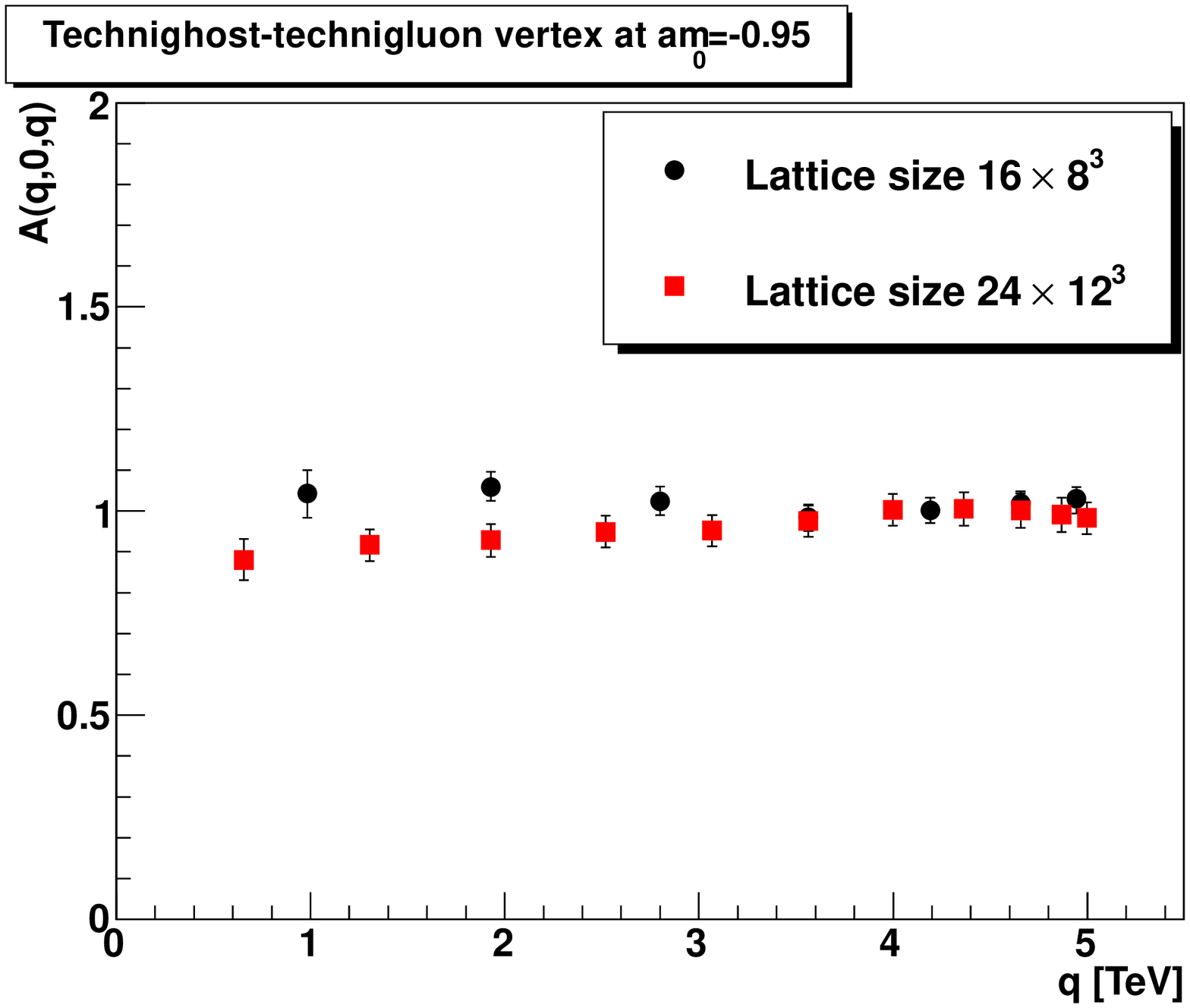}\includegraphics[width=0.5\linewidth]{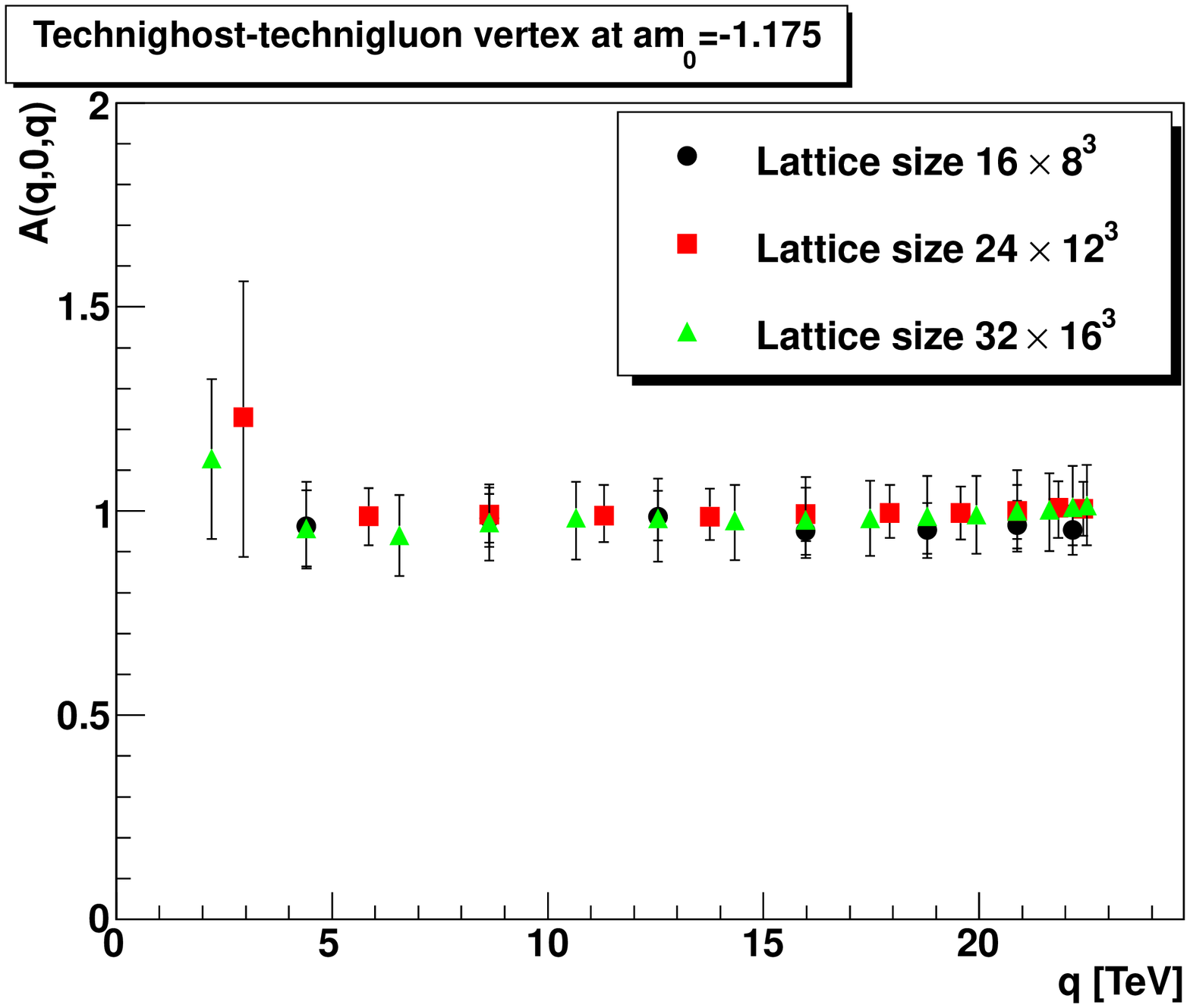}
\caption{\label{fig:ggv-v}The technighost-technigluon vertex dressing function $A$ as a function of the lattice size at otherwise fixed parameters. The left-hand panel shows the results for $am_0=-0.95$, the right-hand panel shows the results for $am_0=-1.175$. Results are renormalized to one at 4.5 TeV for $am_0=-0.95$ and at 21 TeV for $am_0=-1.175$.}
\end{figure}

Though the three-point vertices in Yang-Mills theory have so far shown less finite-volume artifacts \cite{Cucchieri:2008qm,Maas:2007uv}, it is worthwhile to see whether this also applies to the present case. The situation for the technighost-technigluon vertex is depicted in figure \ref{fig:ggv-v}. Within statistical errors, no significant volume dependence is seen, though there might be a small systematic effect at the larger bare techniquark mass.

Furthermore, in Landau gauge the three-point vertices are perturbatively finite. Still, finite renormalizations may be performed, and have been done here. In Yang-Mills theory, the differences in such finite renormalizations for different volumes are below the statistical errors \cite{Cucchieri:2008qm}. Here, they are sizable, in particular closer to zero bare mass they differ by about a third. A possible explanation for this observation is given by the results on a symmetric lattice, as shown in figure \ref{fig:vert-asy}. If this result is reliable, then the vertex shows a variation with the technigluon momentum, but not with the technighost momentum. Since the results on the asymmetric lattice are at zero technigluon momentum, the volume-dependence of the renormalization constant could therefore be just a result of this dependence on the technigluon momentum. Therefore, the overall scale of the technighost-technigluon vertex is completely volume-dependent, and cannot be used as indicative for anything here. However, the dependence on the technighost momentum is still meaningful, and this dependence shows, as stated, no significant volume dependence within statistical errors, though there might exist some.

\begin{figure}
\includegraphics[width=0.5\linewidth]{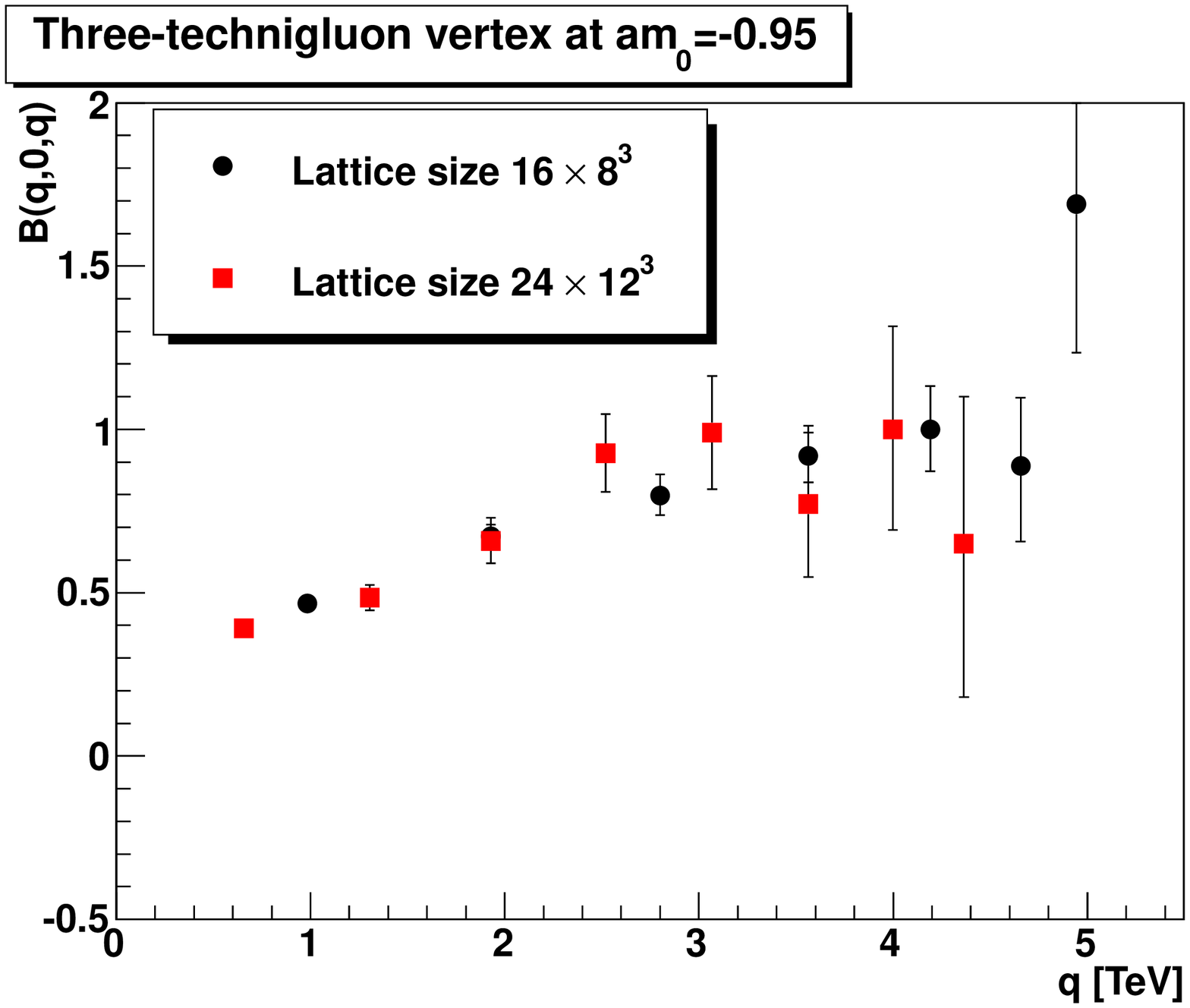}\includegraphics[width=0.5\linewidth]{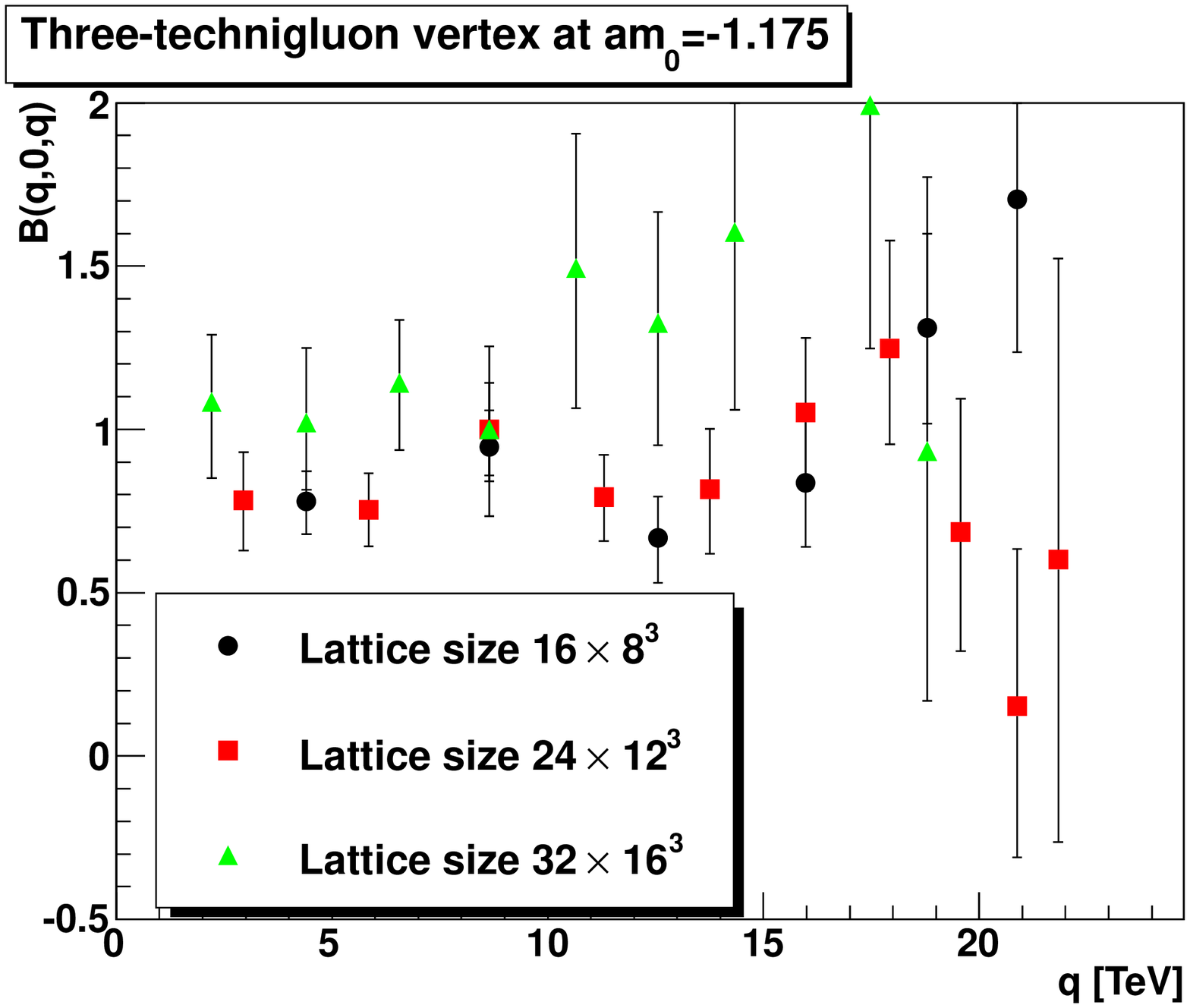}
\caption{\label{fig:g3v-v}The three-technigluon vertex dressing function $B$ as a function of the lattice size at otherwise fixed parameters. The left-hand panel shows the results for $am_0=-0.95$, the right-hand panel shows the results for $am_0=-1.175$. Results are renormalized at 4 TeV for $am_0=-0.95$ and at 8.5 TeV for $am_0=-1.175$ to one.}
\end{figure}

The situation for the three-technigluon vertex is depicted in figure \ref{fig:g3v-v}. Within the large statistical errors no significant volume dependence is found. In this case also no exceptional dependence of the renormalization constants on the volume has been found. However, given the Bose symmetry of this tensor structure of the vertex there are fewer possibilities available for a differing momentum behavior than in case of the technigluon-technighost vertex.

\subsection{Phase dependence}

One of the interesting observations for the current system, which will also play a role when interpreting the results in section \ref{sec:inter}, is the fact that when moving towards the critical value of $\kappa$ at some point a spatial phase transition occurs, i.\ e.\ the center symmetry in spatial directions becomes broken \cite{DelDebbio:2010hx}. The corresponding techniquark bare masses at which this occurs are shown in figure \ref{fig:scale} for two lattice volumes. Results for other lattice volumes can be found in \cite{DelDebbio:2010hx}. In Yang-Mills theory, such a situation is only known when the spatial extension becomes smaller than the inverse finite-temperature phase transition temperature. In analogy, such a breaking is not observed in the temporal direction in the present case, as the temporal extent is much larger than the spatial one.

The influence of such a phase transition on the investigated correlation functions is sizable, in particular when moving far away from the phase transition \cite{Cucchieri:2007ta,Fischer:2010fx,Cucchieri:2010ft}. However, these effects are strongly influenced by finite-volume effects, making their precise analysis complicated \cite{Cucchieri:2007ta,Cucchieri:2010ft}. In contrast, results from the studies of Yang-Mills theories indicate that scalar quantities, like the ghost, or quantities polarized along the direction with unbroken center symmetry, are only weakly influenced \cite{Cucchieri:2007ta,Fischer:2010fx}. Since only these quantities are investigated here, it can be expected that the consequences of the phase transition may not be too severe.

Furthermore, in the present case it is not possible to disentangle the influence of the phase transition fully from the dependence on the volume discussed in section \ref{sec:volume} and on the techniquark mass to be discussed in section \ref{sec:mass}. These two control parameters completely determine the spatial phase of the system. Thus an independent investigation of phase transition effects is not really possible, and their systematic influence can be expected to be included in the combined systematic effects from the asymmetry, the volume, and the mass.

\subsection{Techniquark mass dependence}\label{sec:mass}

The last of the external control parameters is the techniquark bare mass $am_0$, or equivalently the hopping parameter $\kappa$. As discussed in section \ref{sec:scale}, this parameter may also influence the scale. As results are not yet available for a wide range of gauge couplings, which could offset the change of scale, here two different comparisons are made instead. One is without setting the scale, i.\ e.\ with the lattice scale, and one is with the scale set by the techniglueball mass. The prior thus takes into consideration the possibility that the scale setting procedure discussed in section \ref{sec:scale} is not appropriate when taking the limit of zero techniquark mass.

\begin{figure}
\includegraphics[width=0.5\linewidth]{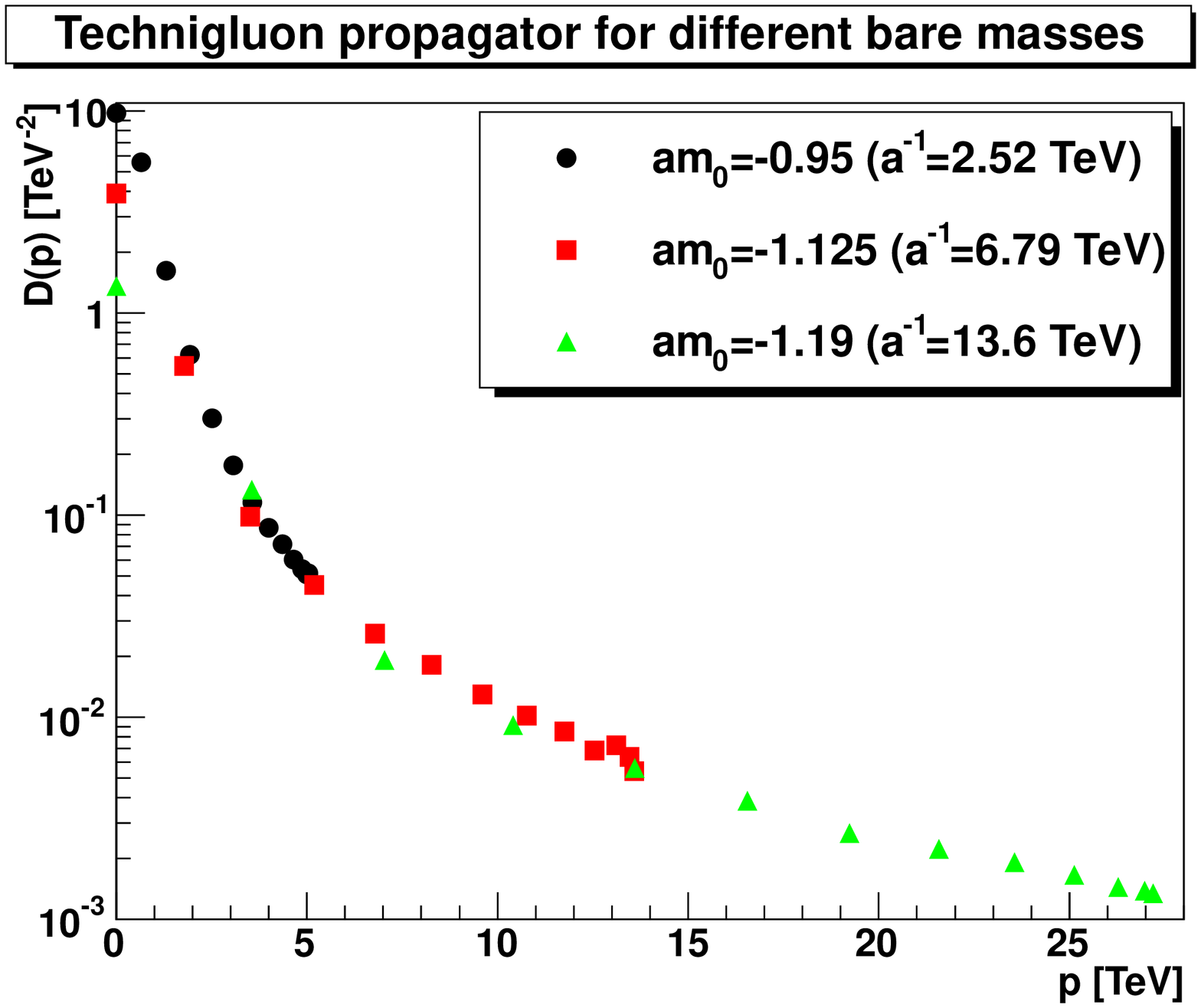}\includegraphics[width=0.5\linewidth]{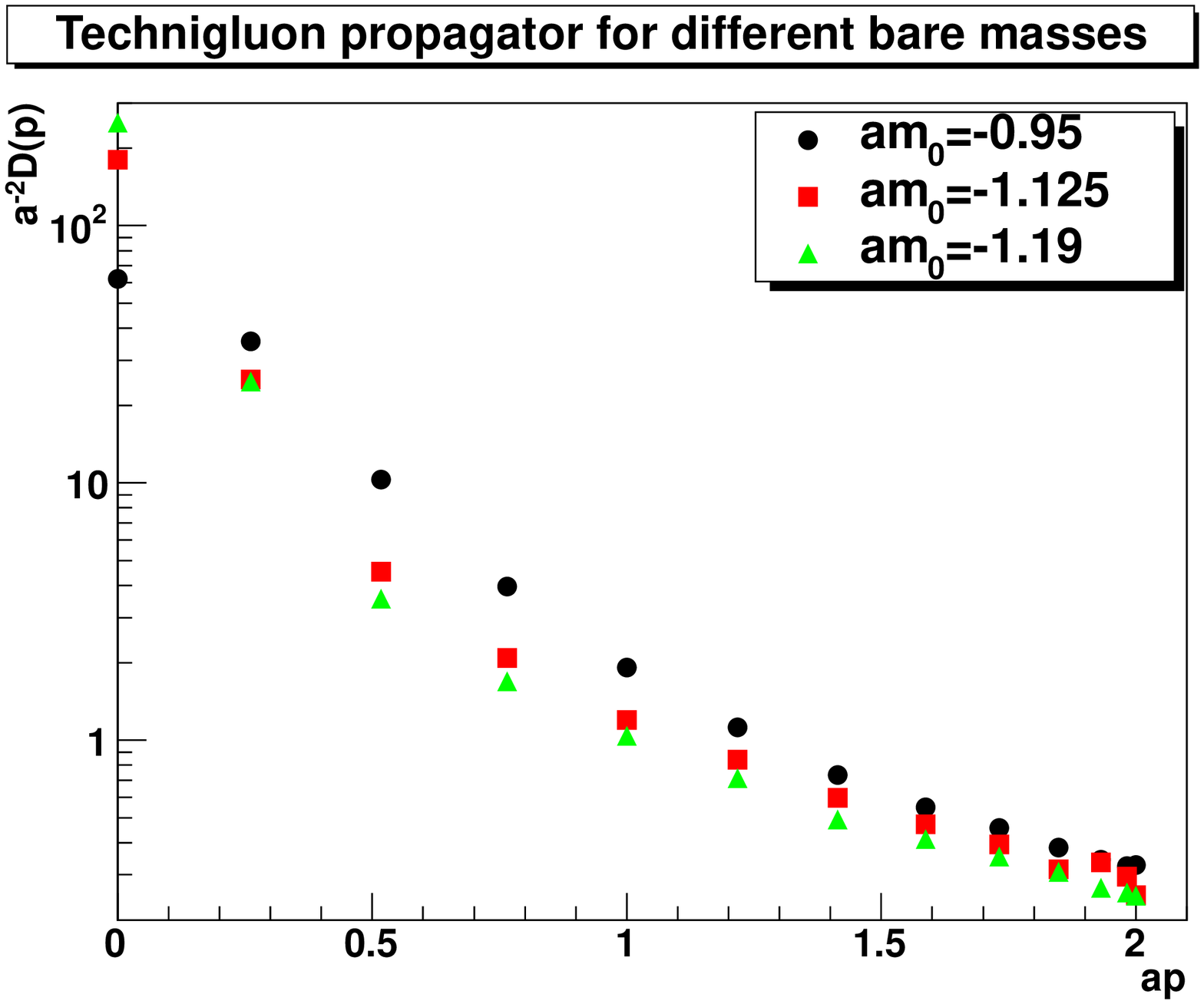}
\caption{\label{fig:gp-a}The technigluon propagator for different bare techniquark masses. The lattice size is $24\times 12^3$. In the left panel the scale from section \ref{sec:scale} has been used. In the right panel the results are given in lattice units. No renormalization has been performed.}
\end{figure}

The first quantity is once more the technigluon propagator, shown in figure \ref{fig:gp-a}. It is immediately visible that in case of the TeV scale the technigluon propagator falls essentially on a universal curve, up to renormalization effects. The remaining differences are then of the same type and size as in the finite-volume case. When the scale is not set, the results do not fall on a universal curve. Moreover, the differences will not be remedied by a momentum-independent renormalization factor, and the differences between the curves is quite large. If the scale were significantly different from the one used here, this would imply that the technigluon propagator is substantially modified when changing the techniquark bare mass.

\begin{figure}
\includegraphics[width=0.5\linewidth]{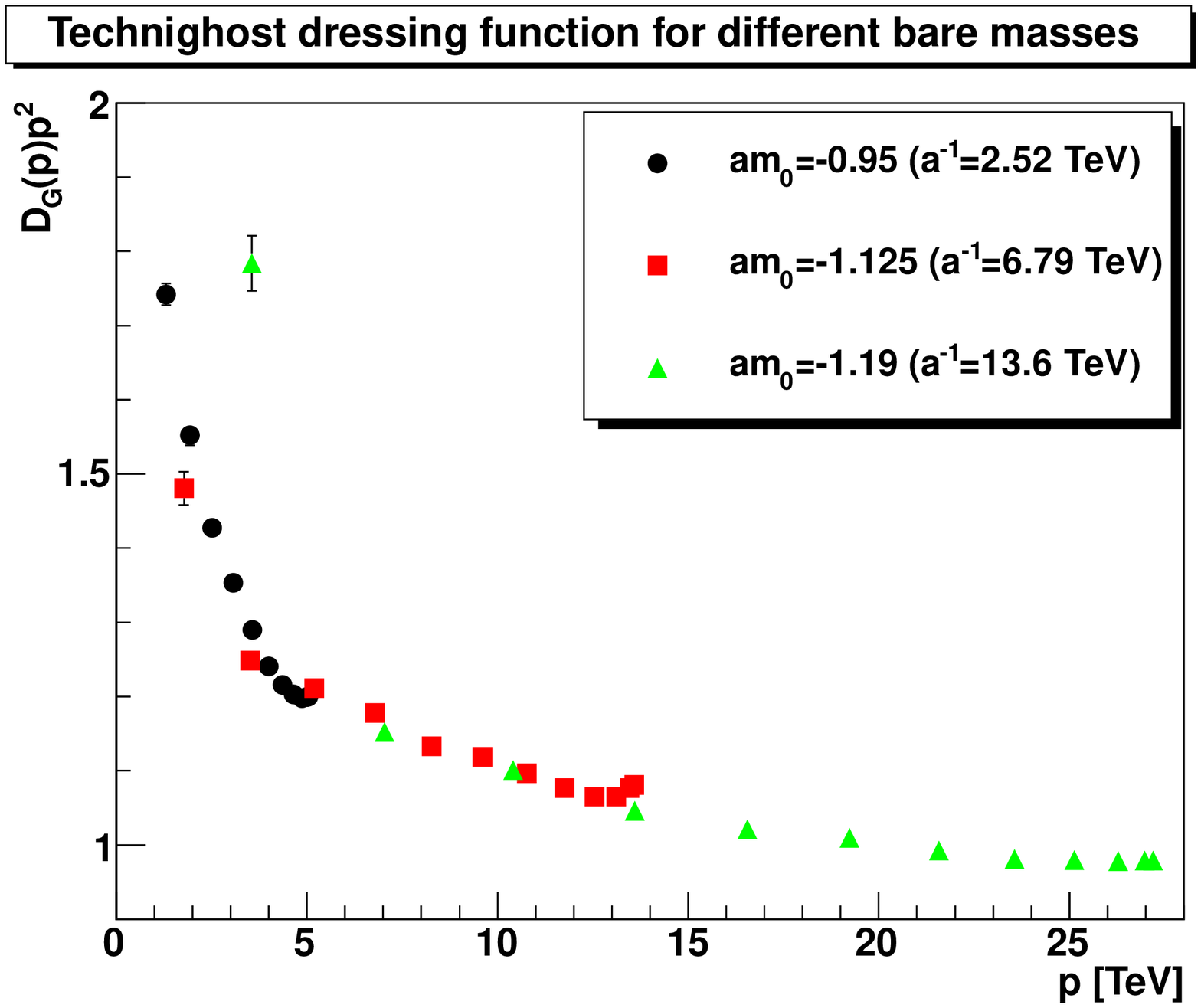}\includegraphics[width=0.5\linewidth]{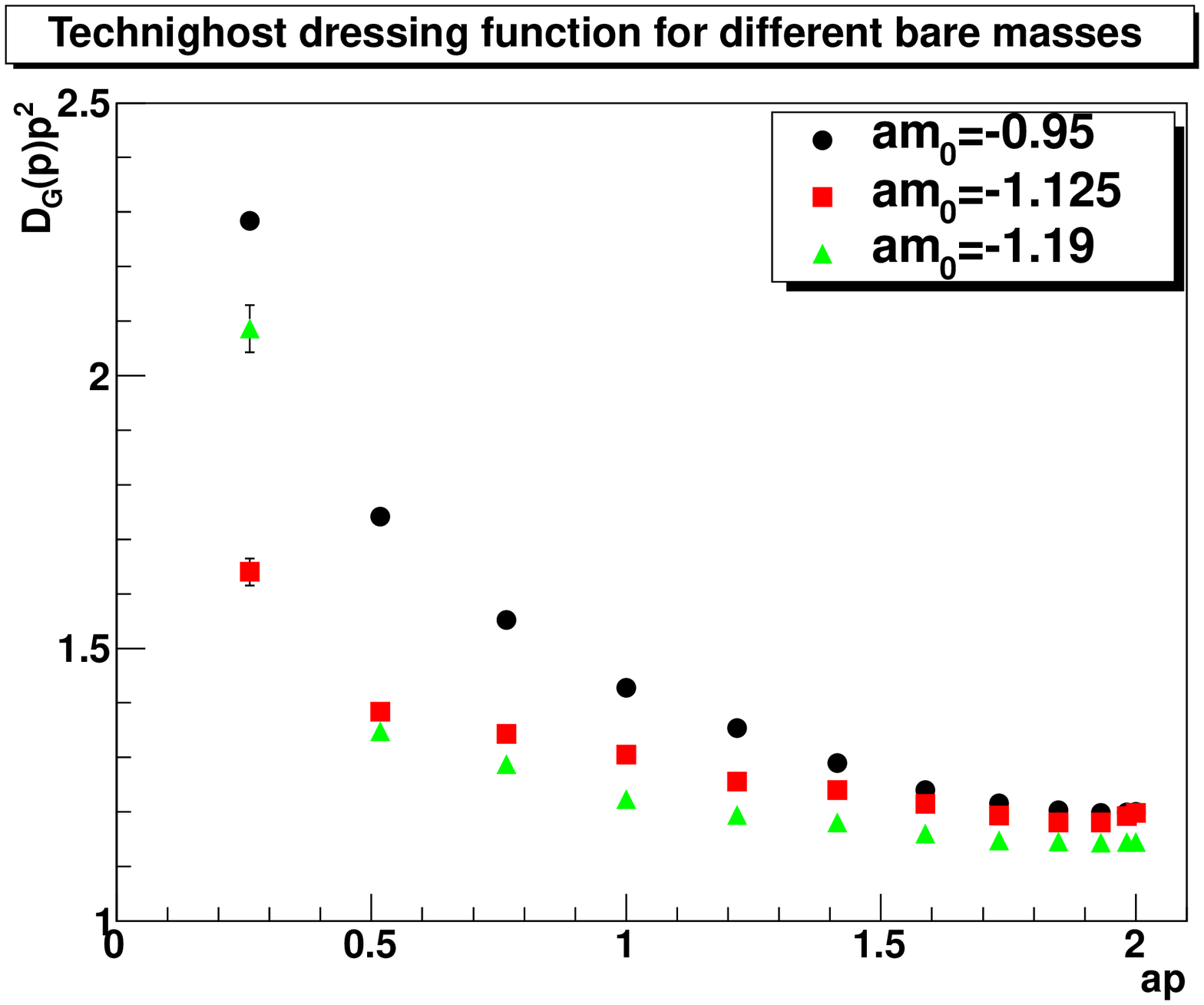}
\caption{\label{fig:ghp-a}The technighost dressing function for different bare techniquark masses. The lattice size is $24\times 12^3$. In the left panel the scale from section \ref{sec:scale} has been used. In the right panel the results are given in lattice units. In the left panel renormalization has been performed.}
\end{figure}

The results for the technighost dressing function are shown in figure \ref{fig:ghp-a}. The situation is similar to the technigluon case\footnote{Note that the light bending up of the ghost propagator at the largest momenta is actually a systematic artifact of low statistics of the numerical method employed, and vanishes for sufficiently large statistics \cite{Cucchieri:2006tf}.}, though in the present case it is necessary to renormalize the dressing function to obtain results, in the version with scale, which lie on a universal curve. Also, the artifacts are somewhat larger than in the finite-volume case, showing that the influence of the techniquark mass is likely not entirely related to the scale. On the other hand, in the scale-free case there is no hint of a universal behavior, implying that if another scale is used the technighost dressing function drastically depends on the techniquark mass.

\begin{figure}
\includegraphics[width=0.5\linewidth]{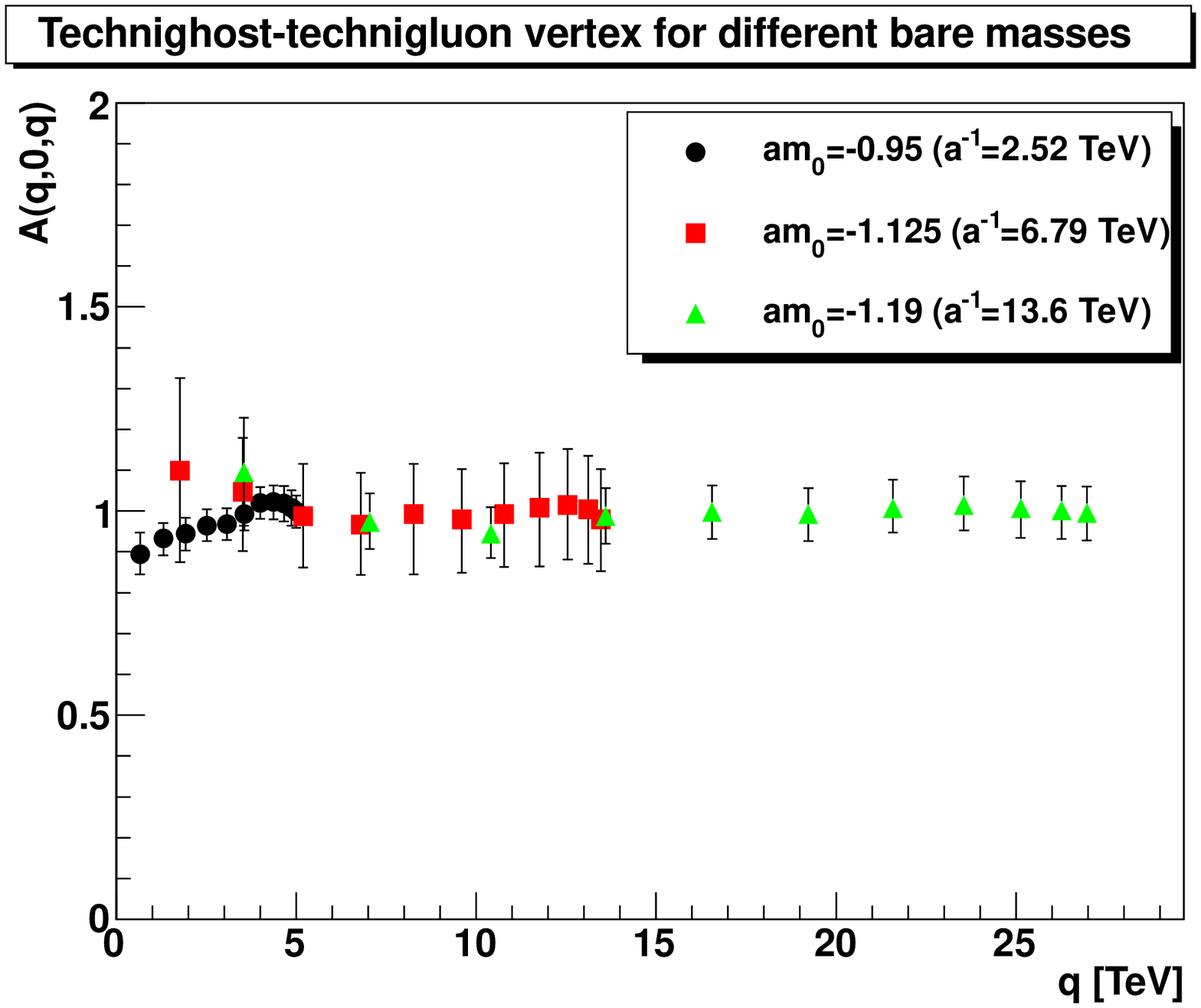}\includegraphics[width=0.5\linewidth]{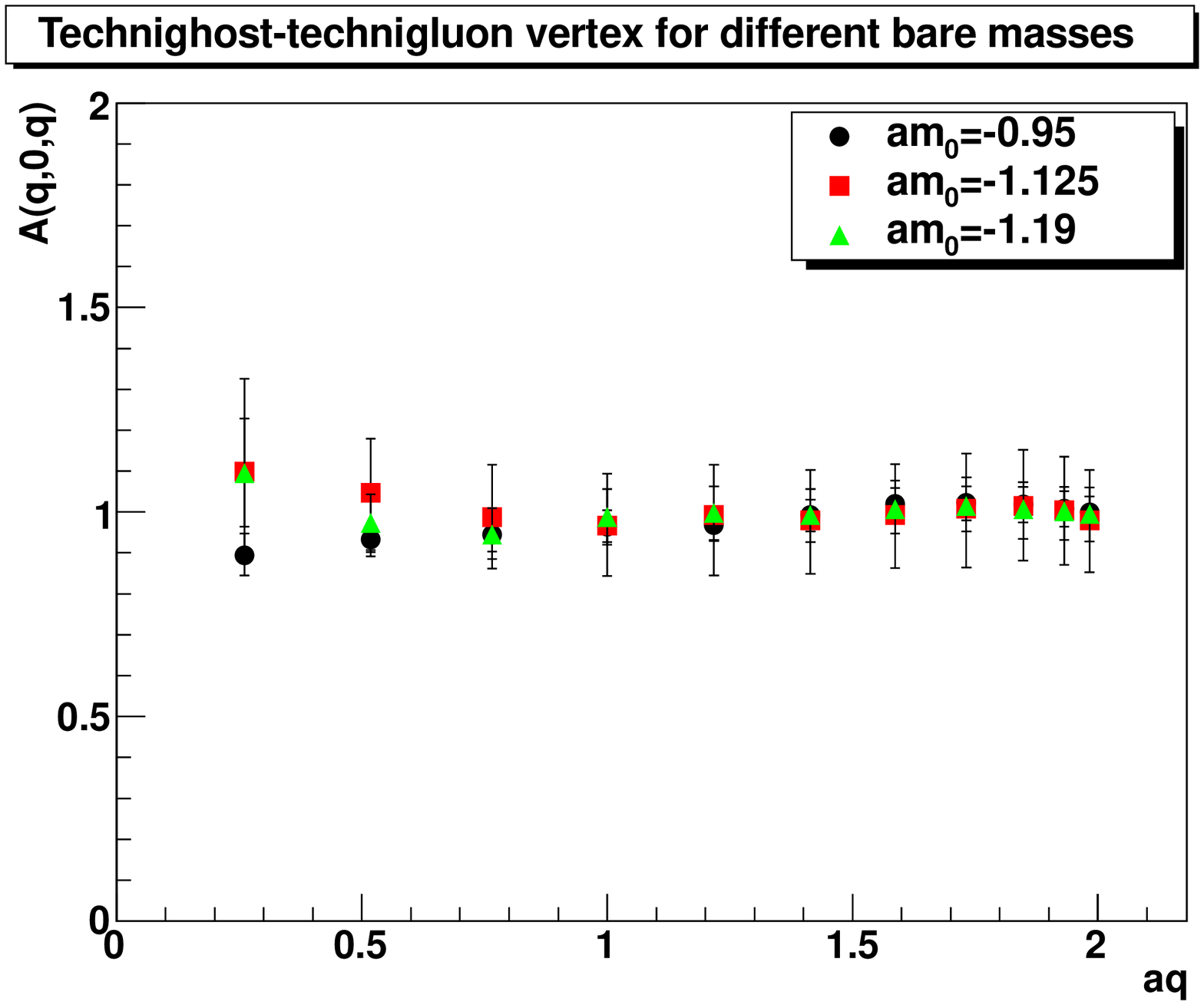}
\caption{\label{fig:ggv-a}The technighost-technigluon vertex dressing function for different bare techniquark masses. The lattice size is $24\times 12^3$. In the left panel the scale from section \ref{sec:scale} has been used. In the right panel the results are given in lattice units. Renormalization has been performed in both cases.}
\end{figure}

The results for the technighost-technigluon vertex are shown in figure \ref{fig:ggv-a}. For both cases, within statistical errors, the results fall on a universal curve. Given that the result is an almost momentum-independent, dimensionless quantity, this is not too surprising. Thus, the technighost-technigluon vertex does not show any pronounced dependency on $\kappa$.

\begin{figure}
\includegraphics[width=0.5\linewidth]{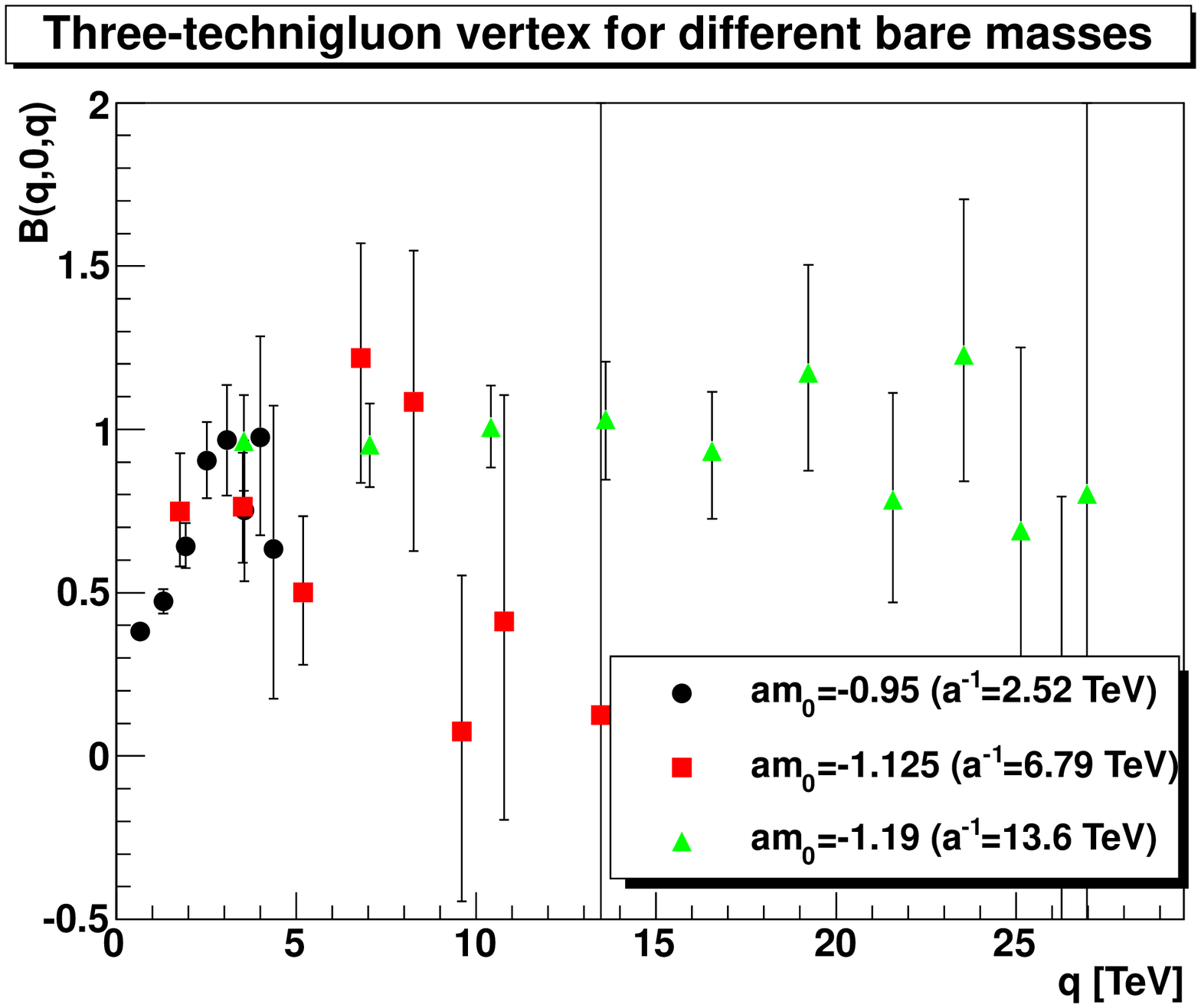}\includegraphics[width=0.5\linewidth]{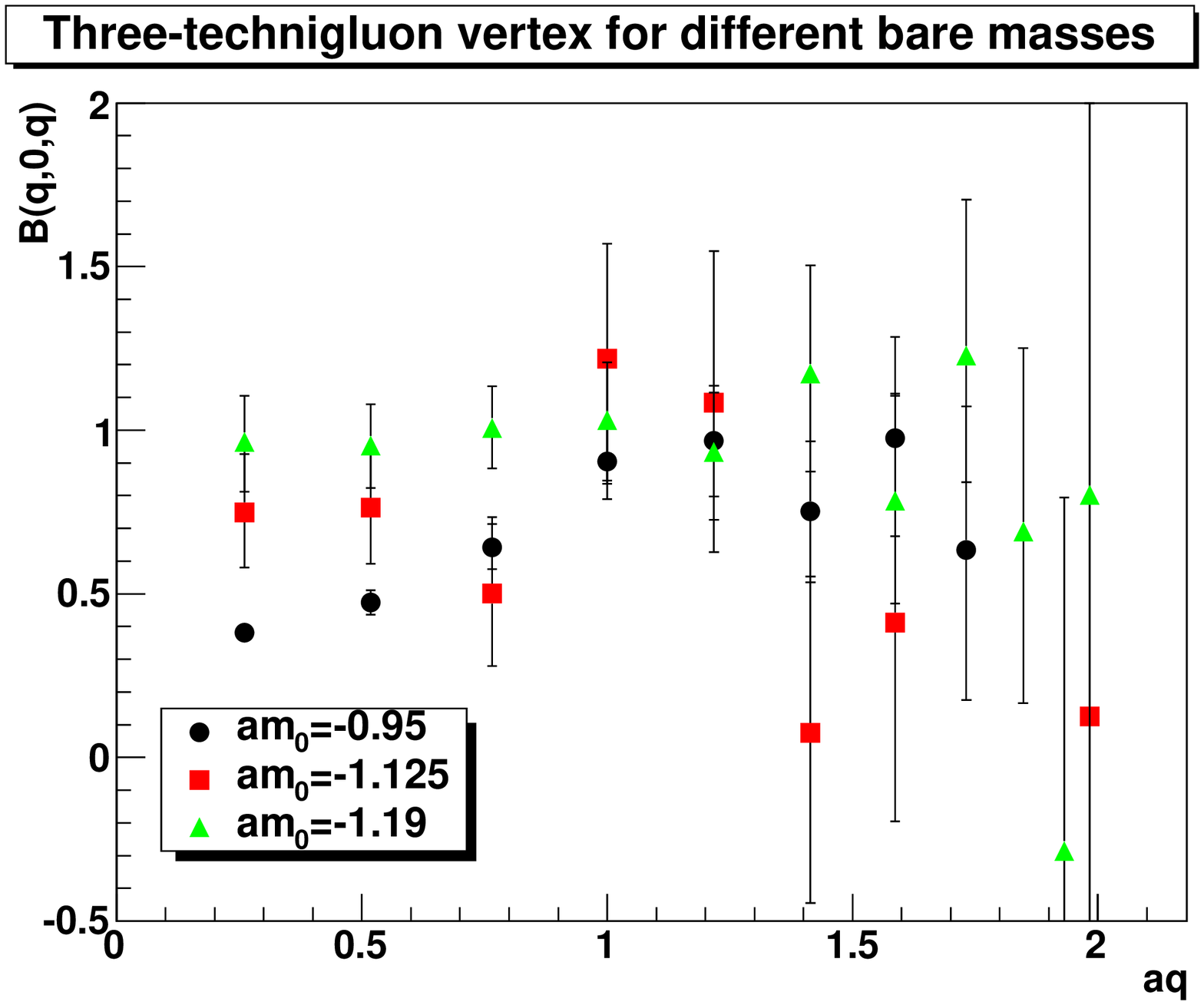}
\caption{\label{fig:g3v-a}The three-technigluon vertex dressing function for different bare techniquark masses. The lattice size is $24\times 12^3$. In the left panel the scale from section \ref{sec:scale} has been used. In the right panel the results are given in lattice units. Renormalization has been performed in both cases.}
\end{figure}

Finally, the results for the three-technigluon vertex are shown in figure \ref{fig:g3v-a}. In contrast to the technighost-technigluon vertex it shows quite a dependence on the techniquark mass, which also manifests itself in the dependency on which scale is chosen. Again, as for the propagators, within the huge statistical errors the curves fall on a universal one after renormalization when the TeV-scale is used. However, the renormalization coefficients turn out to be not a monotonous function of the bare techniquark mass. If the lattice scale is used instead, there is once more a dependence on the techniquark mass.

In total, thus, the dependence on the techniquark mass is the strongest effect, and much stronger than in the case of QCD \cite{Kamleh:2007ud}. A more detailed discussion of the possible implications of this result will be given in the interpretation in section \ref{sec:inter} after the presentation of the final results in section \ref{sec:res}.

\subsection{Gauge dependence}\label{sec:gauge}

The last systematic effect to be treated is not a lattice artifact, but a field-theoretical effect due to the non-perturbative Gribov-Singer ambiguity \cite{Gribov:1977wm,Singer:1978dk}. This ambiguity implies that the specification of the gauge as Landau gauge is not a sufficient definition of a gauge beyond perturbation theory \cite{Gribov:1977wm}, and has to be resolved using additional non-local conditions \cite{Singer:1978dk}. In the main part of the text this is done using the so-called minimal Landau gauge. Of course, since the studied correlation functions are gauge-dependent, it should be investigated to which extent the results depend on this choice. In the following, choices will be studied which are based on a selection of Gribov copies different from the minimal Landau gauge, the absolute Landau gauge and two Landau-$B$ gauges, the min$B$ and the max$B$ gauge, see \cite{Maas:2009se,Maas:2008ri,Zwanziger:1993dh} for their definitions. At least for small lattices at finite cutoff these yield quite different results for the correlation functions from minimal Landau gauge \cite{Maas:2009se,Bornyakov:2008yx}. The precise situation in the continuum is not yet fully resolved, see \cite{Maas:2009se,Maas:2010wb,Fischer:2008uz} for an introduction to the current state of affairs. However, for the present purpose this is not of relevance.

To study these different gauges, it is necessary to have access to at least part of the gauge orbit after imposing the perturbative gauge condition, i.\ e., to more than one Gribov copy of each residual gauge orbit. Unfortunately, with current algorithms it is not possible to guarantee finding all Gribov copies. Nonetheless, since for the present purpose only an indication of effects is desired, a subset of copies should be sufficient. Therefore, for each configuration up to twenty attempts are made to find Gribov copies. The actual number found is then at most twenty, see for details \cite{Maas:2009se,Cucchieri:1997dx}.

\begin{figure}
\includegraphics[width=\linewidth]{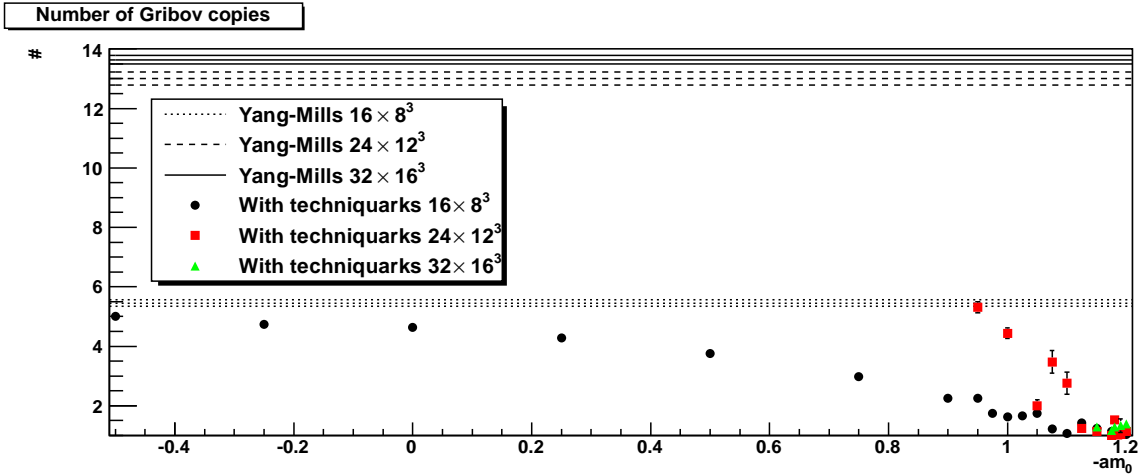}
\caption{\label{fig:gc}The number of Gribov copies as a function of the techniquark mass (symbols) compared to Yang-Mills theory (lines) for various lattice volumes at fixed $\beta=2.25$.}
\end{figure}

The first interesting observation made is that the number of Gribov copies found is significantly dependent on $\kappa$. At a $\kappa$ far away from the critical one, the number found is of about the same order as for Yang-Mills theory with the same lattice parameters. When approaching the critical $\kappa$, this number quickly decreases, to almost only one Gribov copy found. This is illustrated in figure \ref{fig:gc}. Such a decrease has so far been observed either when in Yang-Mills theory the volume is reduced \cite{Maas:2009se,Maas:2008ri}, or when moving from the confinement to the Higgs phase in a Yang-Mills-Higgs theory \cite{Maas:2010nc}. However, in the present case the number of Gribov copies also increases with increasing lattice volume at fixed $\kappa$, as in Yang-Mills theory.

\begin{figure}
\includegraphics[width=0.5\linewidth]{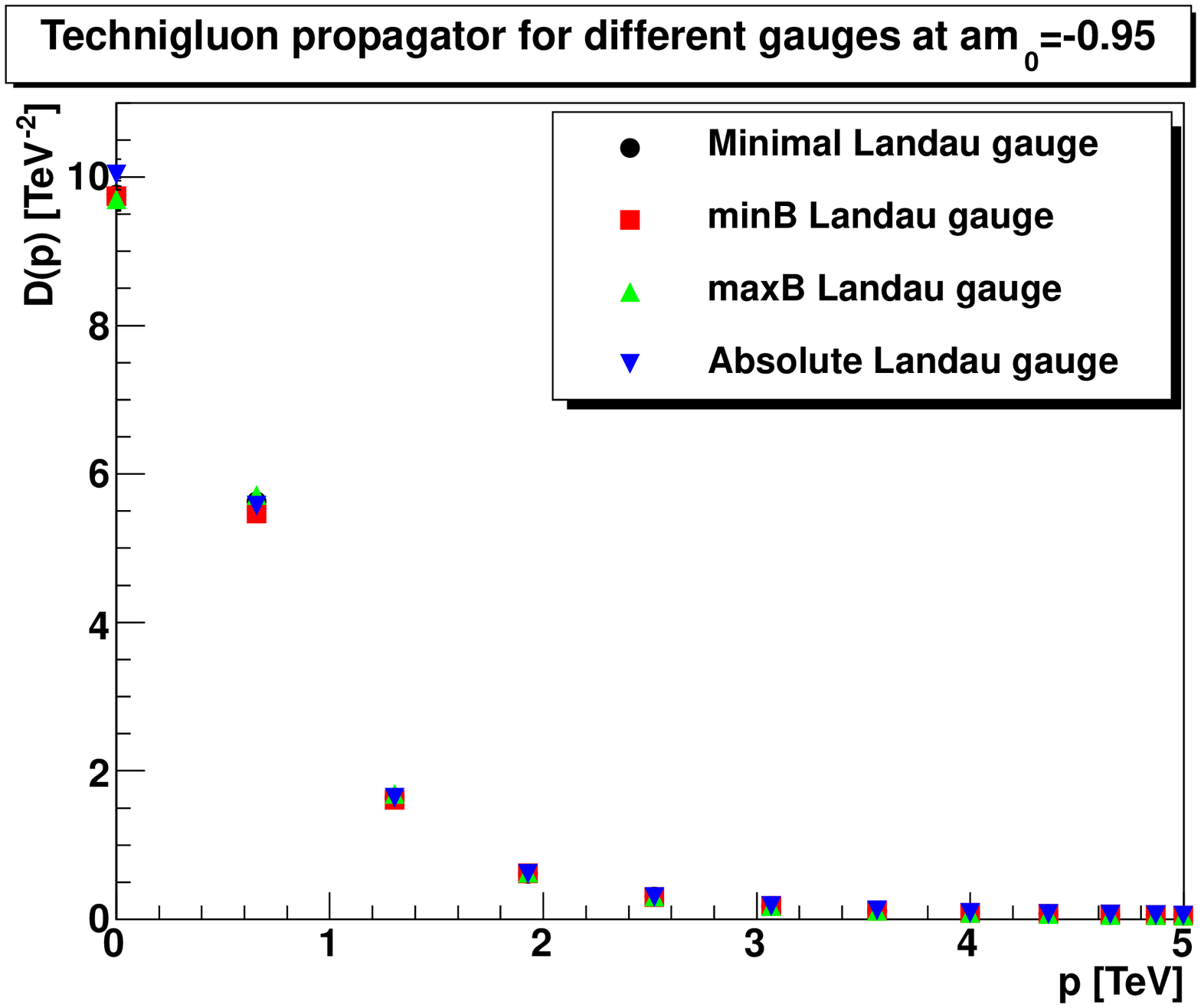}\includegraphics[width=0.5\linewidth]{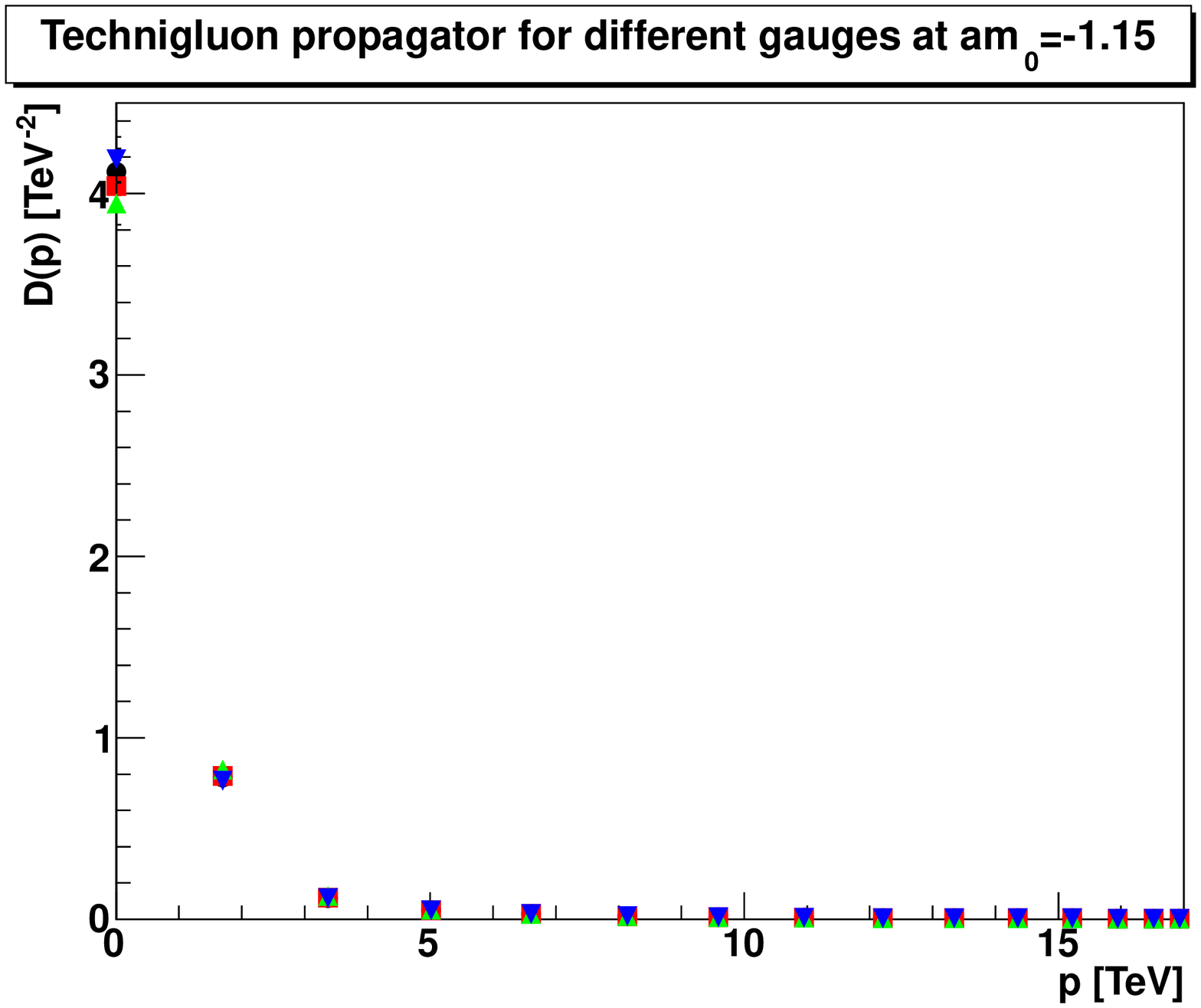}\\
\includegraphics[width=0.5\linewidth]{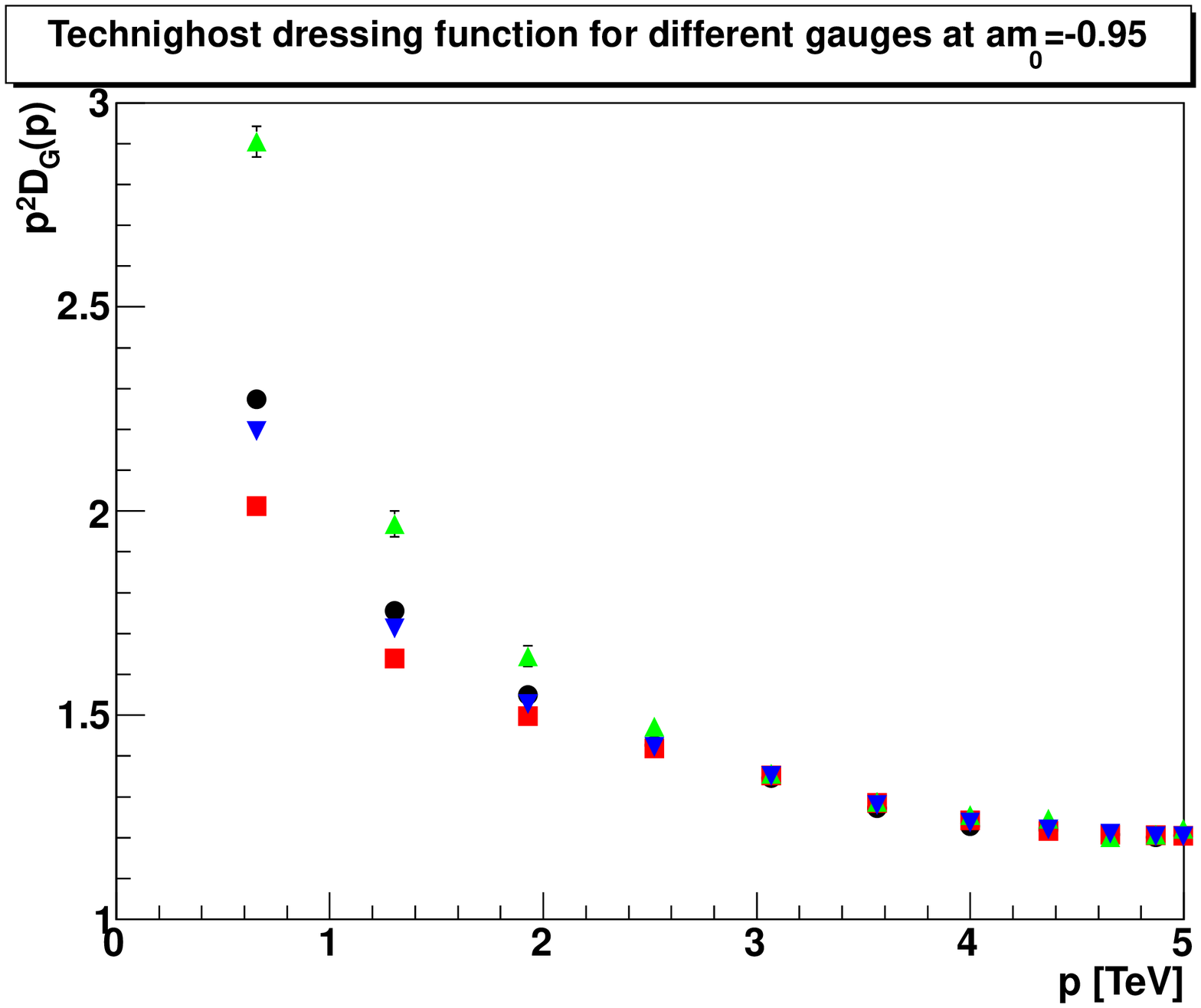}\includegraphics[width=0.5\linewidth]{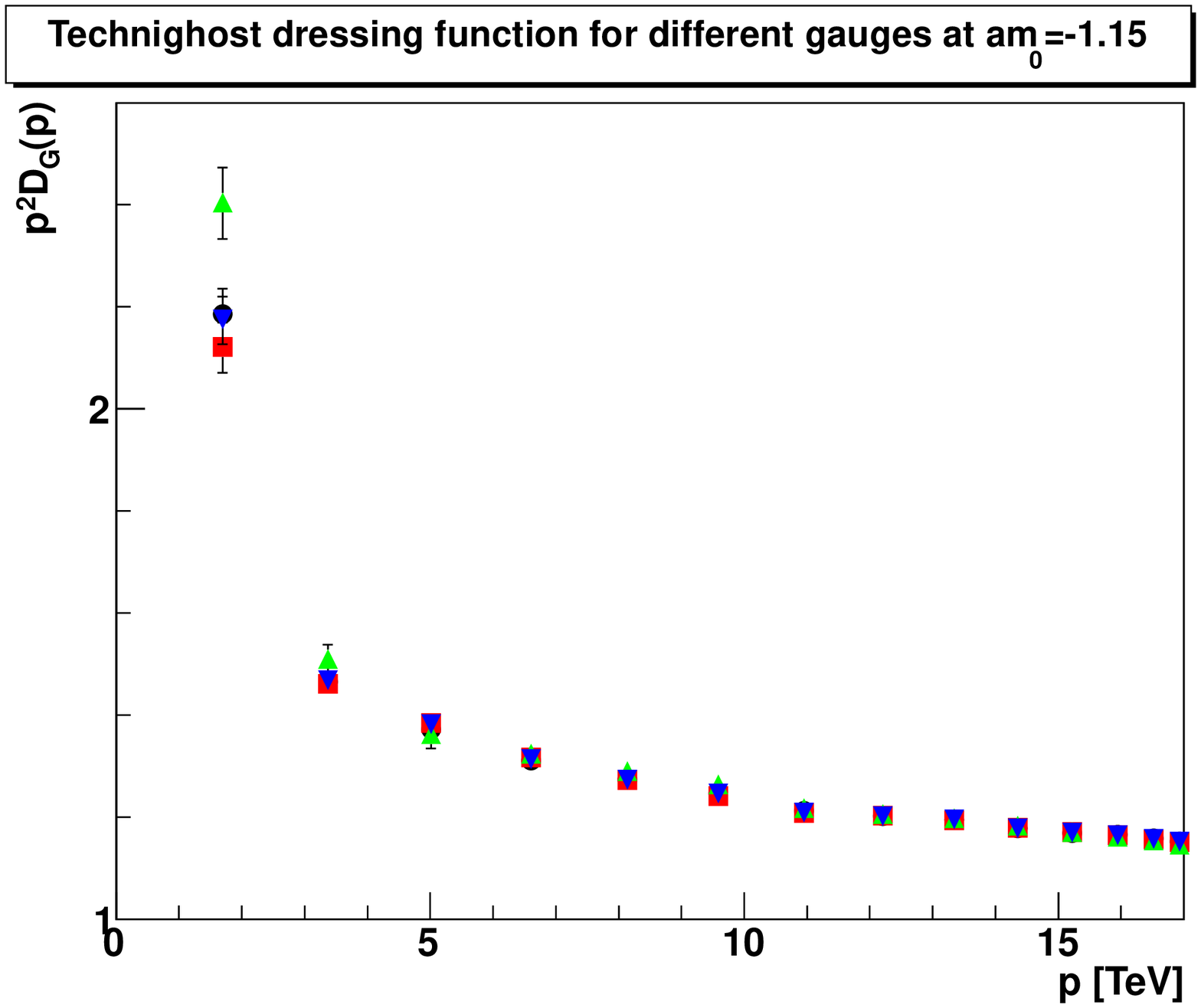}
\caption{\label{fig:gauges}The technigluon propagator (top panels) and the technighost dressing function (bottom panels) for different non-perturbative gauge choices. The left panels are for a $24\times 12^{3}$ lattice at $am_0=-0.95$ and the right panels for a $32\times 16^3$ lattice at $am_0=-1.15$.}
\end{figure}

The influence of the choice of gauge is illustrated for the technigluon propagator and the technighost dressing function for two different values of the bare techniquark mass in figure \ref{fig:gauges}. The variability found is in agreement with what is found for Yang-Mills theory at the same number of Gribov copies \cite{Maas:2009se}. In particular, the technigluon propagator is essentially unaffected by the gauge choice, while the technighost shows quite a dependency at small momenta. There does not appear to be any influence of the techniquark mass other than changing the number of Gribov copies. Hence, as in the case of Yang-Mills-Higgs systems in the confinement phase \cite{Maas:2010nc}, the sensitivity to the non-perturbative gauge choice is not significantly affected by the presence of the techniquarks.

\subsection{Summary of systematic effects}

Summarizing, the results emphasize that the dominant errors from lattice artifacts are located at zero momentum. In particular, the results on the technigluon at zero momentum cannot be trusted, and are therefore discarded in the section on the results. Furthermore, the lowest non-zero momentum point still shows sizable effects on the order of a few ten percent, and thus should be treated with caution. Finally, the techniquark mass has a strong influence on the results, which will be of significant importance when interpreting the results in section \ref{sec:inter}.

Of course, one should be wary that this can at best be a lower limit on the systematic uncertainties. The experience with Yang-Mills theory \cite{Cucchieri:2008fc,Cucchieri:2007rg,Bornyakov:2009ug,Bogolubsky:2009dc,Fischer:2007pf,Sternbeck:2008mv,Cucchieri:2006za,Cucchieri:2007ta} implies that much more detailed investigations are necessary to obtain at least a reasonable measure of control over systematic errors. In particular, to get control over volume artifacts will require much larger, symmetric lattice volumes. In addition, the unexpected behavior of the technigluon-technighost vertex also suggests the necessity of investigating symmetric lattices for a wide range of lattice parameters and techniquark mass.

\section{Results}\label{sec:res}

In the following the results will be presented, and compared to Yang-Mills theory. For this purpose, results will be used from the largest lattice size available with sufficient statistics for a given value of the techniquark mass. For the Yang-Mills case the comparison will be made with two different $\beta$ values each, depending on the parameters of the largest lattices available. The comparison will always be made twice, once with the scale set as discussed in section \ref{sec:scale}, and once with the lattice scale. Also 3d plots will be shown with the correlation functions as a function of momenta and the bare techniquark mass. However, in these cases different volumes are mixed, and finite-volume effects will be more relevant at small momenta.

\begin{figure}
\includegraphics[width=0.5\linewidth]{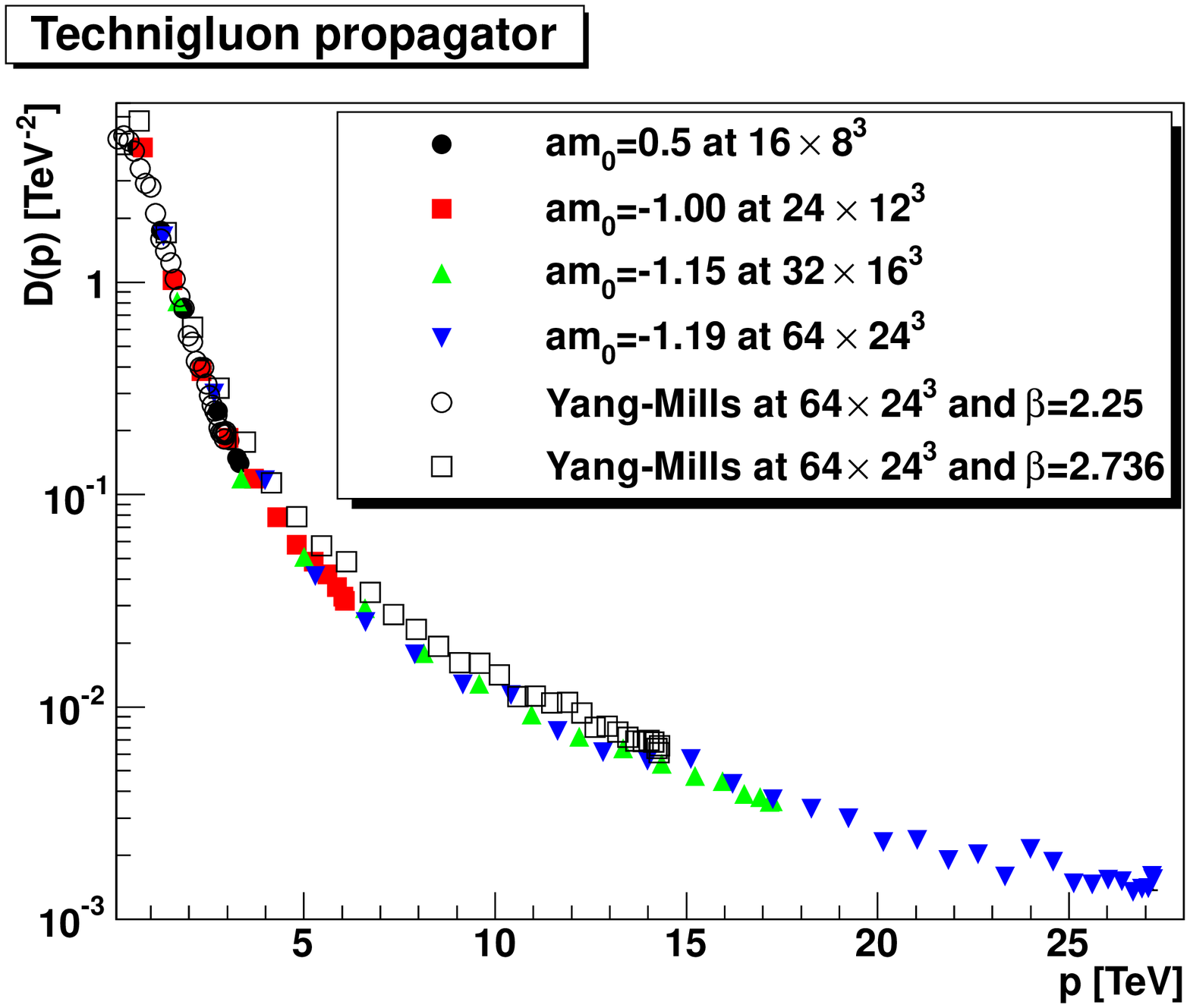}\includegraphics[width=0.5\linewidth]{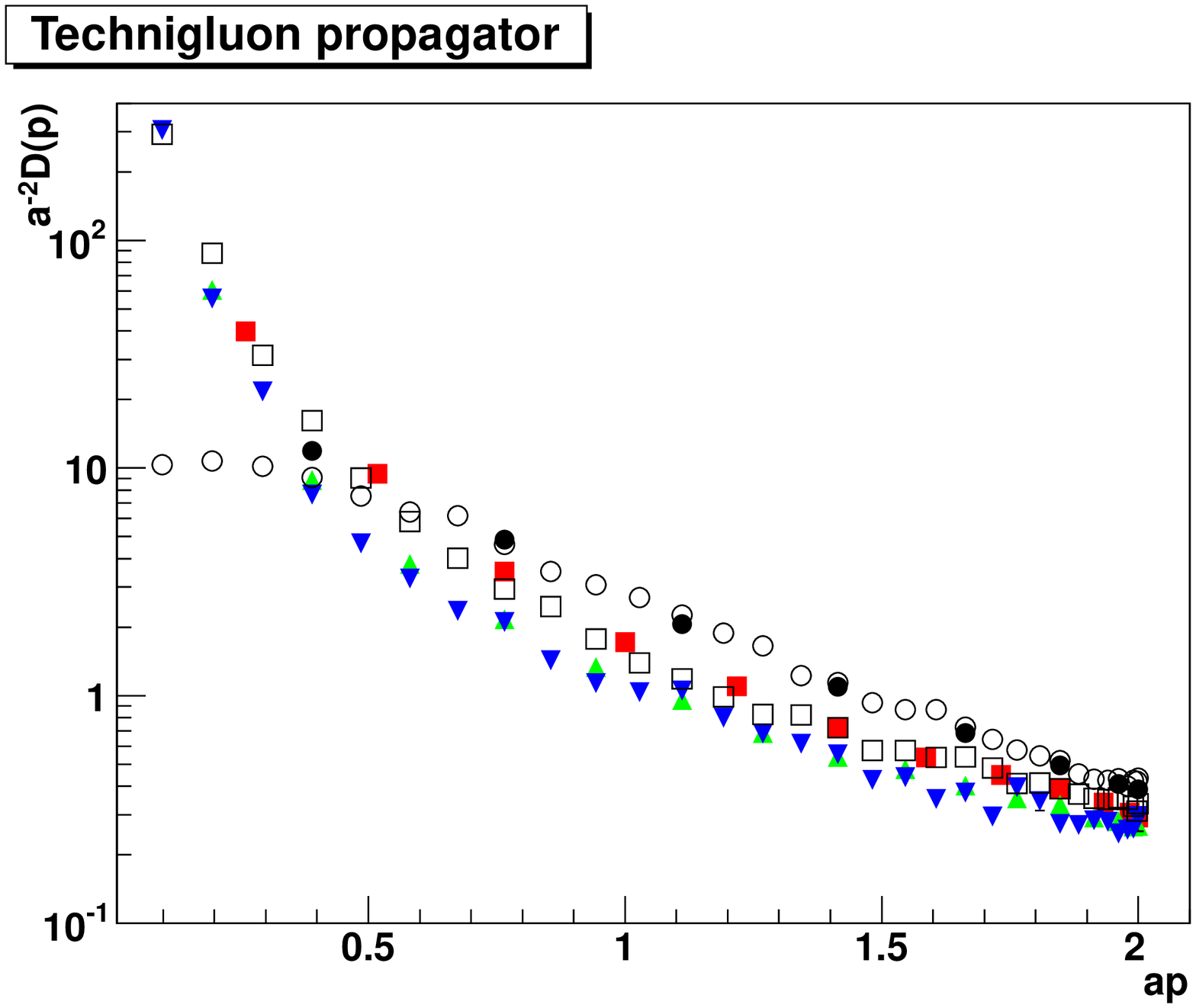}\\
\includegraphics[width=0.5\linewidth]{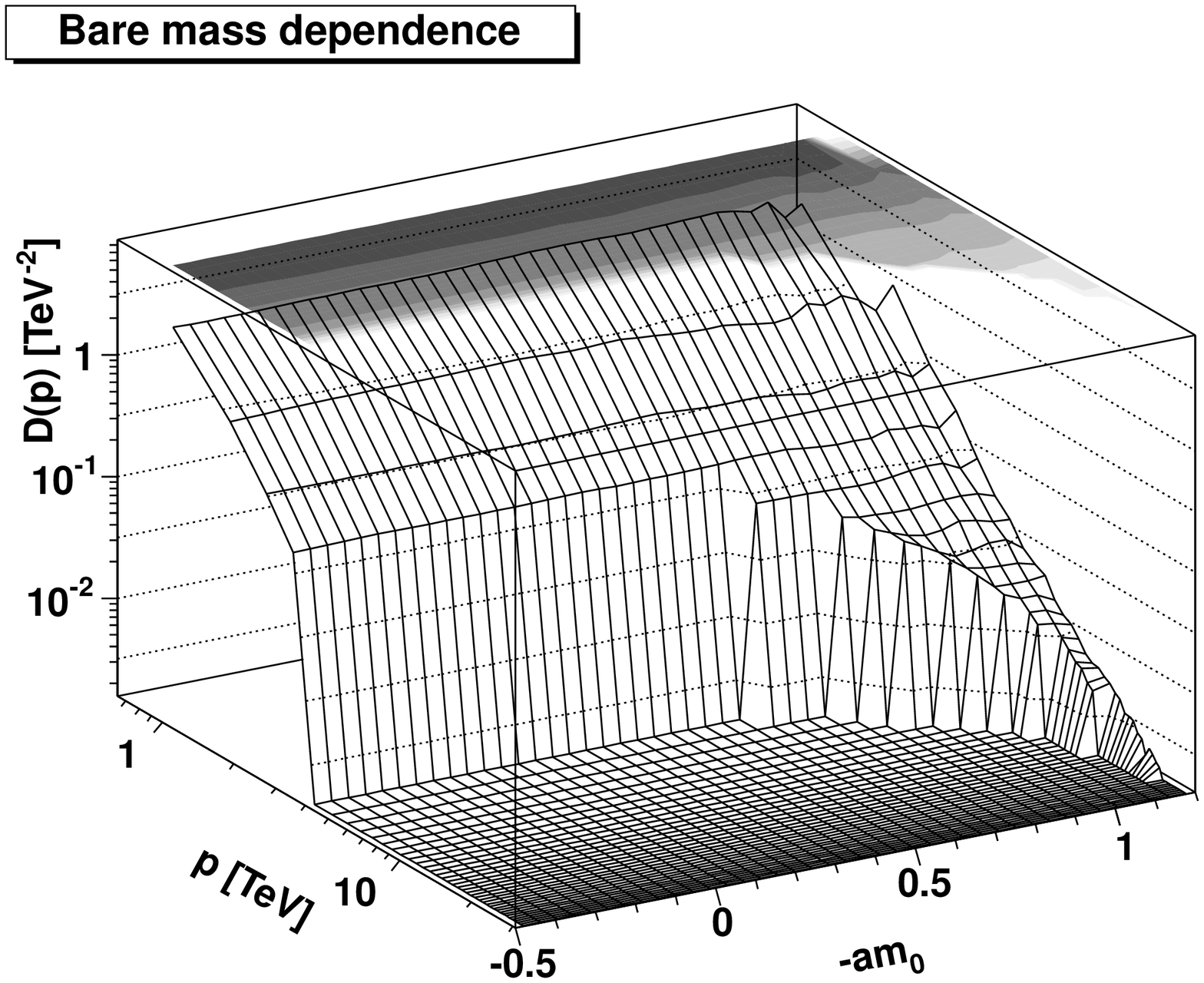}\includegraphics[width=0.5\linewidth]{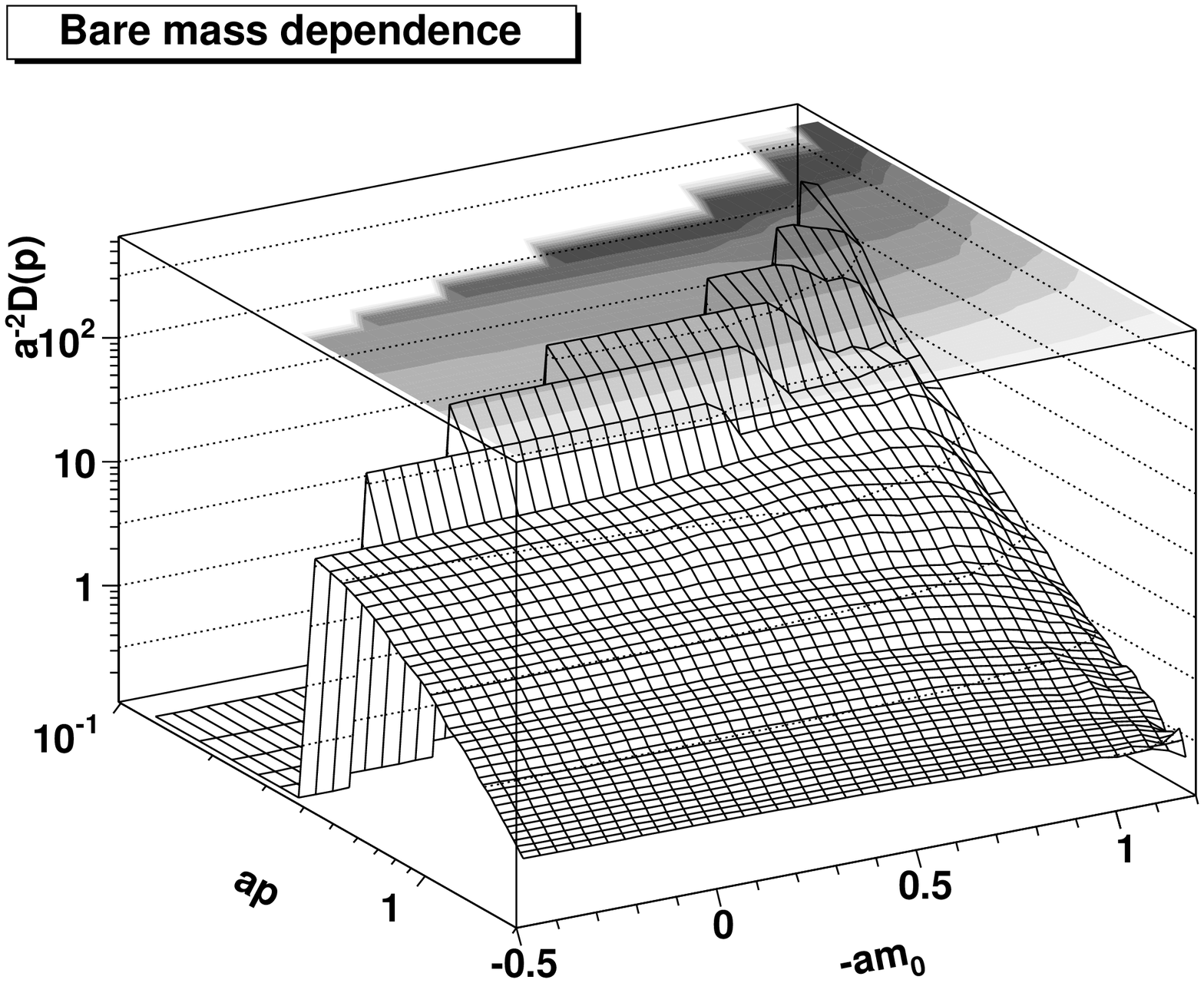}
\caption{\label{fig:gp}The technigluon propagator with (top-left panel) and without scale (top-right panel) for different bare masses. The lower two panels show the propagator as a function of both momenta and bare mass, again with (bottom-left panel) and without (bottom-right panel) scale. In the three-dimensional plots zero values indicate that no measurement is available at that combination of parameters. No renormalization has been performed.}
\end{figure}

The results for the technigluon propagator are shown in figure \ref{fig:gp}. When using the TeV scale, all results essentially fall on a universal curve, up to minor deviations. This curve furthermore agrees rather well with the Yang-Mills curve, within the systematic uncertainties expected for the employed scale. In particular, the smaller bare masses agree rather well with those of Yang-Mills theory at a larger $\beta$ value. In contrast, for the lattice scale the results do not fall on a universal curve, and also show a stronger deviation from the Yang-Mills case. However, in total the dependency on the bare mass is not too pronounced.

\begin{figure}
\includegraphics[width=0.5\linewidth]{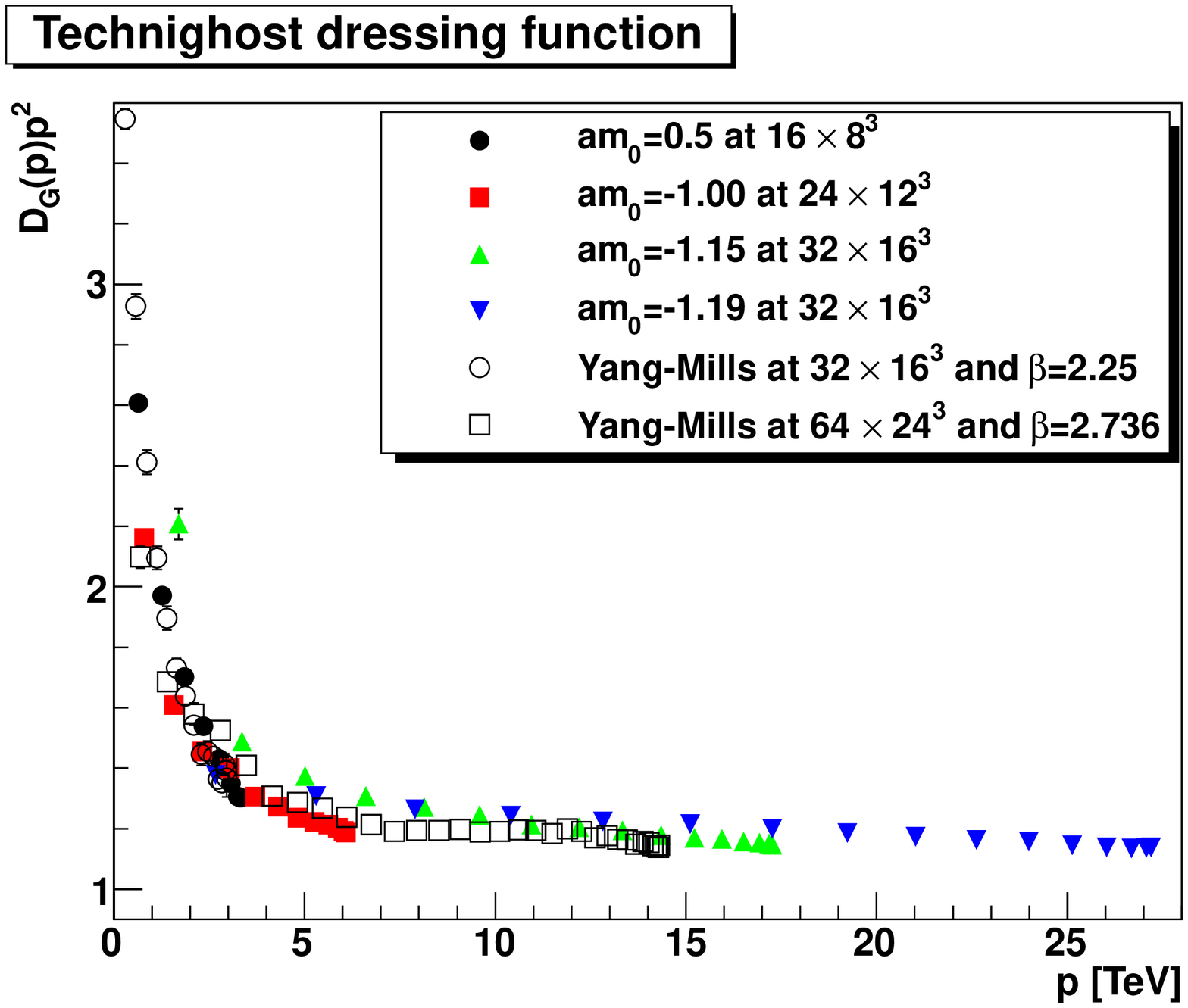}\includegraphics[width=0.5\linewidth]{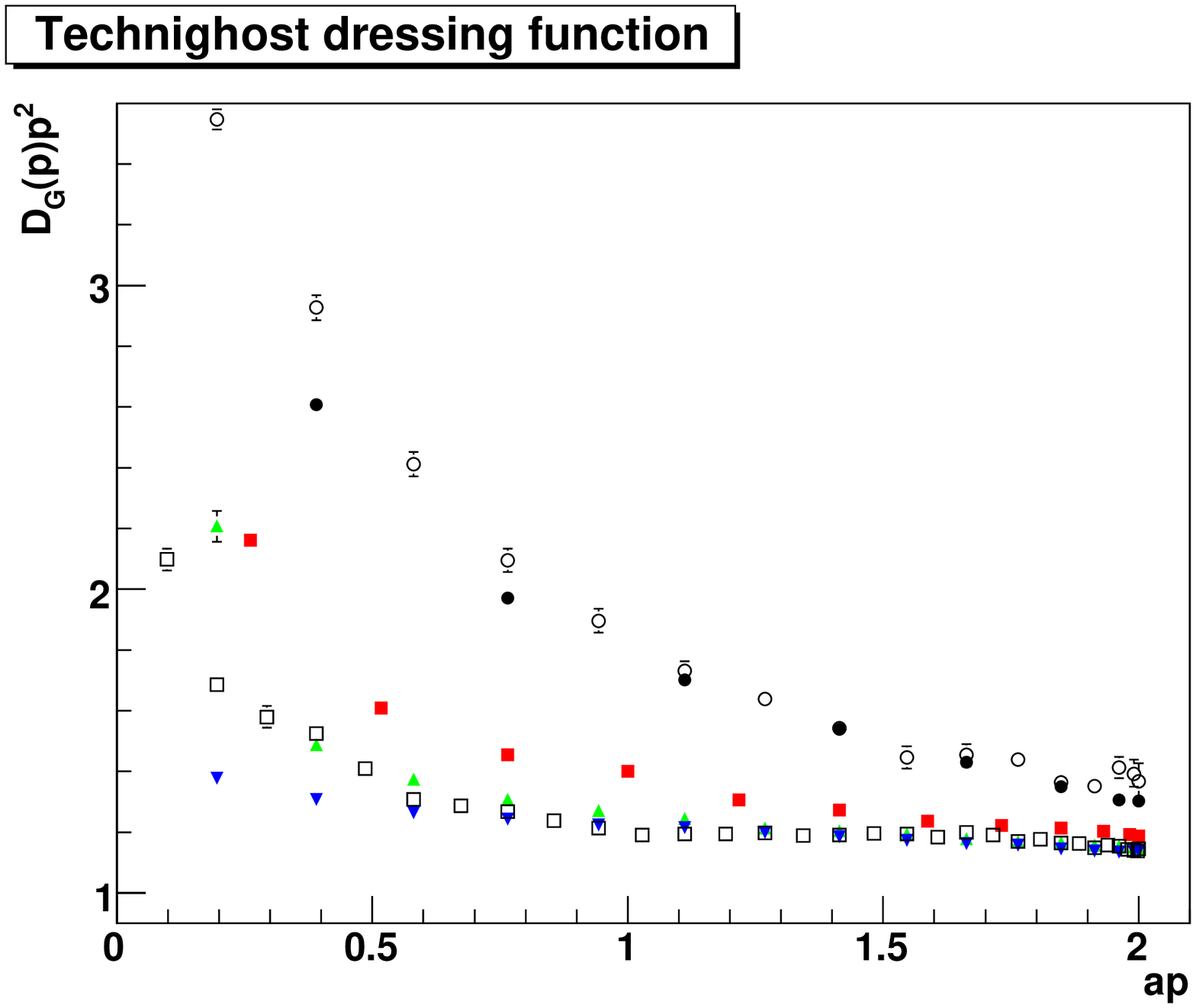}\\
\includegraphics[width=0.5\linewidth]{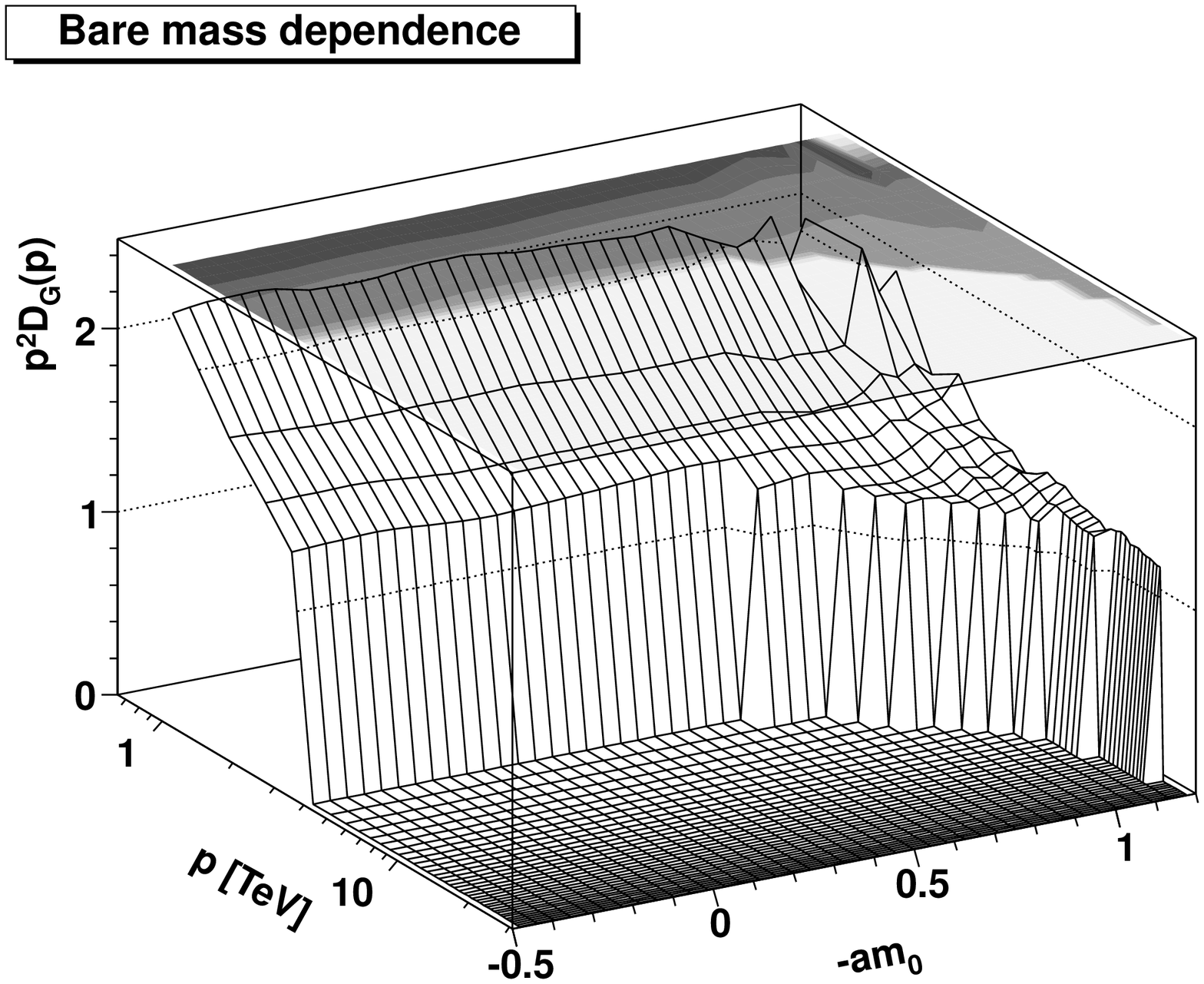}\includegraphics[width=0.5\linewidth]{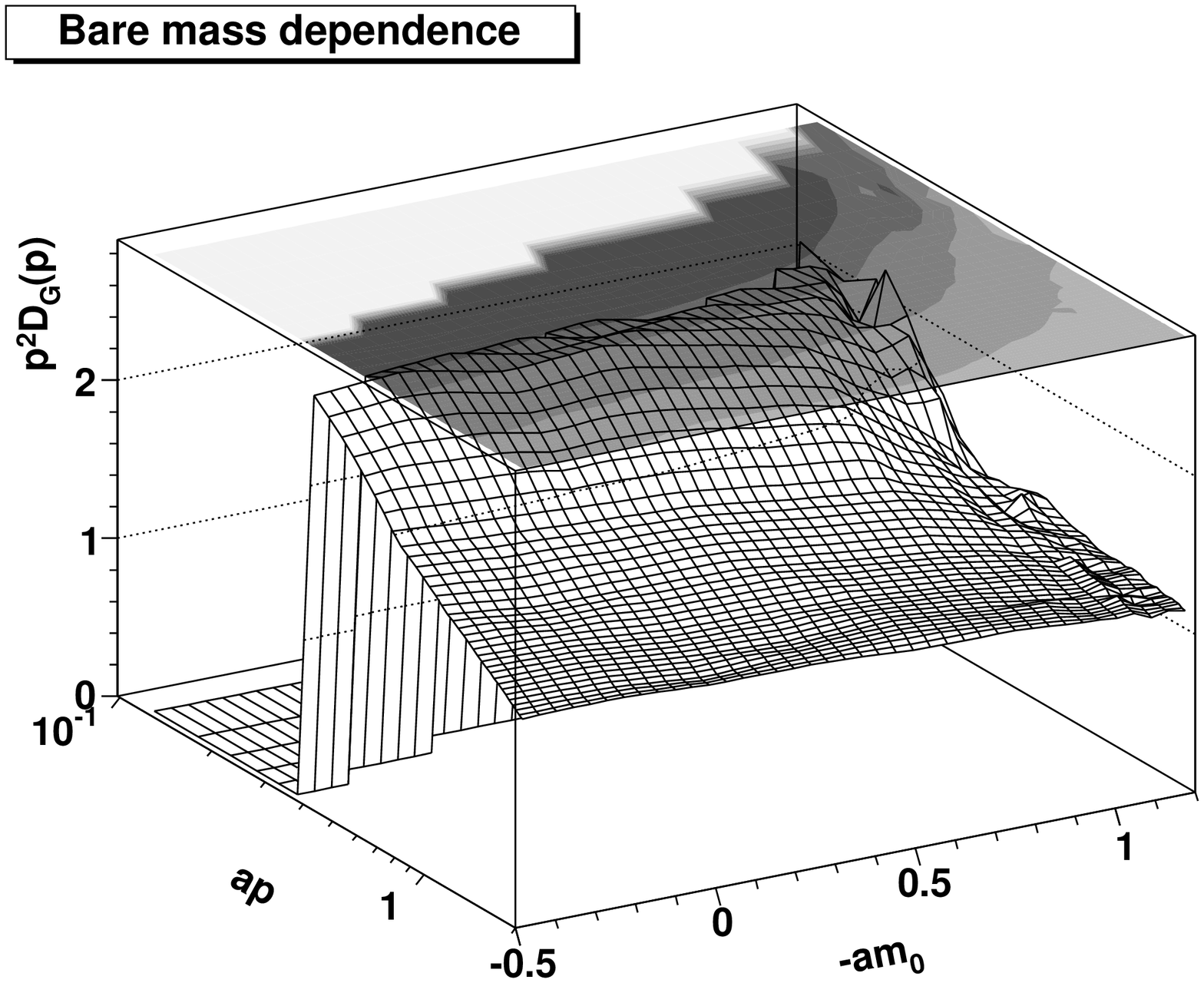}
\caption{\label{fig:ghp}The technighost dressing function with (top-left panel) and without scale (top-right panel) for different bare masses. The lower two panels show the dressing function as a function of both momenta and the bare mass, again with (bottom-left panel) and without (bottom-right panel) scale. In the three-dimensional plots zero values indicate that no measurement is available at that combination of parameters. No renormalization has been performed. Fluctuations at large $\kappa$ in the three-dimensional plots are a statistical artifact, which is somewhat amplified in the left panel by the interpolation.}
\end{figure}

The second quantity is the technighost dressing function, shown in figure \ref{fig:ghp}. Its bare mass dependence is somewhat stronger than in the case of the technigluon propagator. If the scale from section \ref{sec:scale} is used, the results fall almost on a universal curve. The trend is similar to the case of Yang-Mills theory, though in detail it is different. In particular, the infrared steepness is not a simple function of the bare mass, though this influence is minor. On the other hand, in lattice units the technighost dressing function becomes flatter and flatter as a function of the bare mass. The behavior is similar to the one of Yang-Mills theory if $\beta$ is increased.

\begin{figure}
\includegraphics[width=0.5\linewidth]{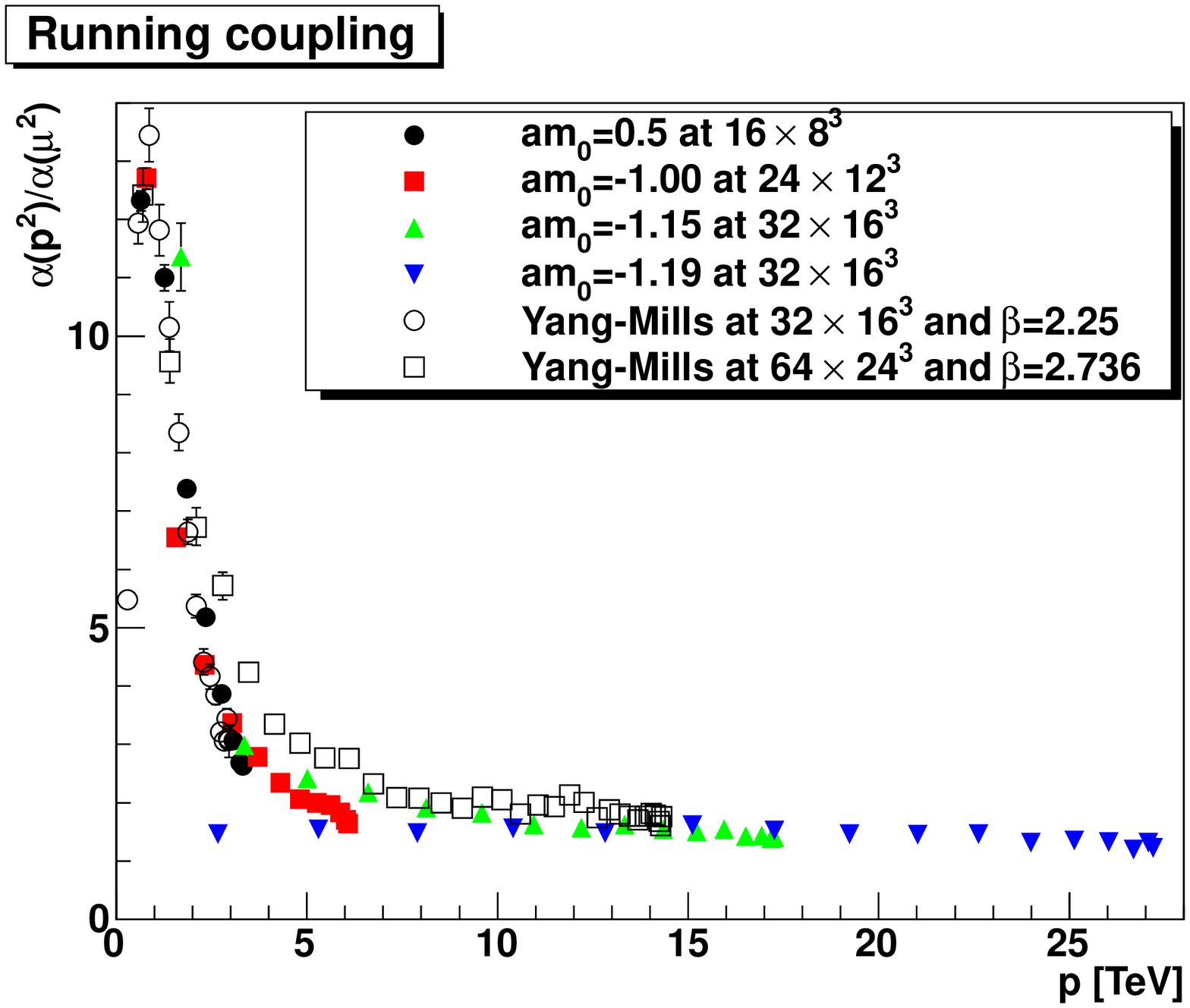}\includegraphics[width=0.5\linewidth]{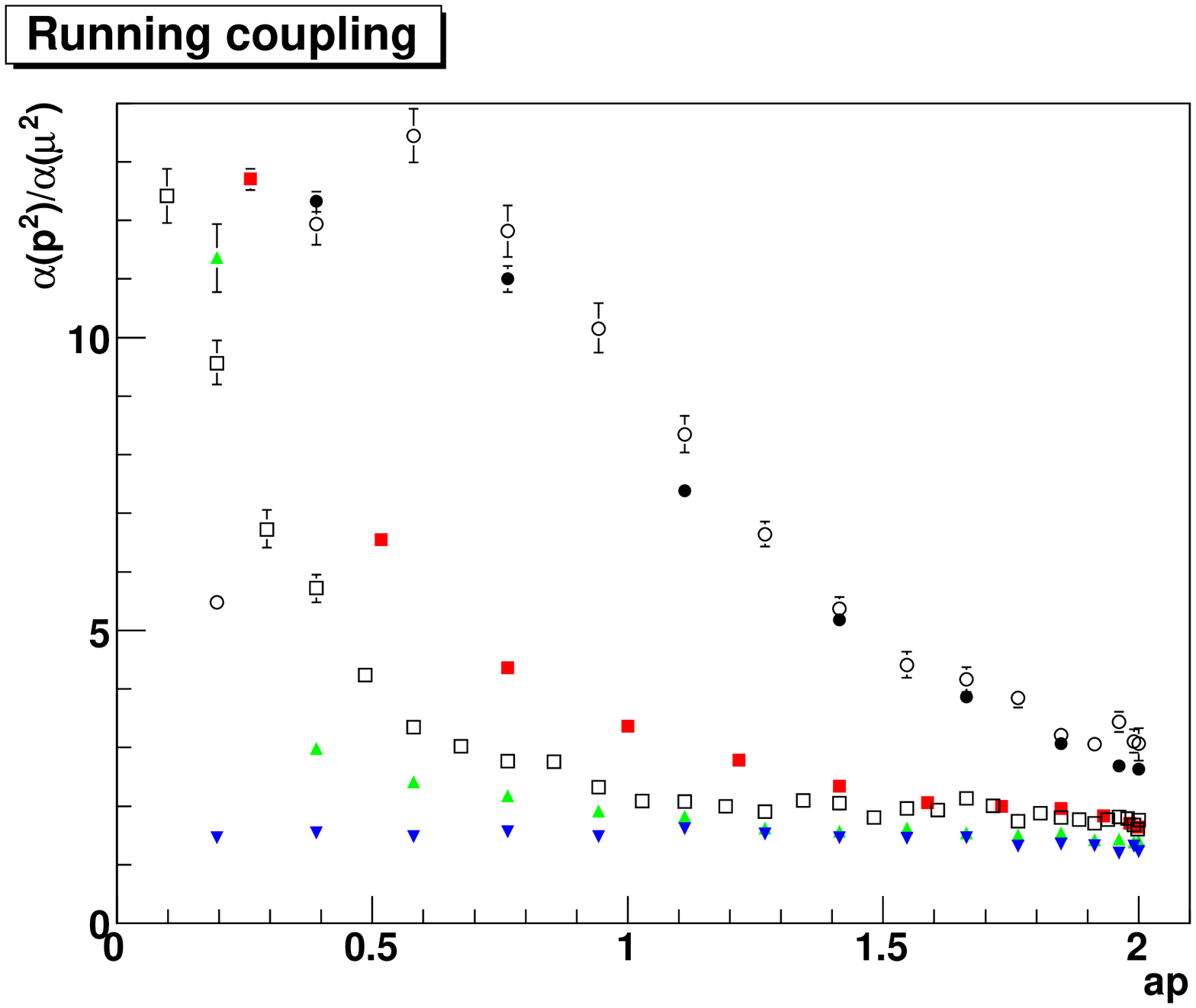}\\
\caption{\label{fig:alpha}The running coupling \pref{alpha} with (left panel) and without scale (right panel) for different bare masses.}
\end{figure}

These modifications amplify themselves if the running coupling \pref{alpha} is investigated, since the propagators enter them as a product. Thus deviations from Yang-Mills theory should be more simple to identify. This comparison is performed in figure \ref{fig:alpha}. The result is rather interesting. Again, there is a tendency for the results to collapse on a common curve when using the scale of section \ref{sec:scale}. The results do not directly follow the Yang-Mills curves, though there is still quite some resemblance to the behavior. In case of the lattice scale, the results show a more and more momentum-independent behavior, which again is to some extent in analogy to Yang-Mills theory. Unfortunately, the statistical accuracy is yet insufficient to determine the $\beta$-function to test its distinct shape \cite{Sannino:2009za}, as is possible in Yang-Mills theory \cite{Maas:2010gi}.

\begin{figure}
\includegraphics[width=0.5\linewidth]{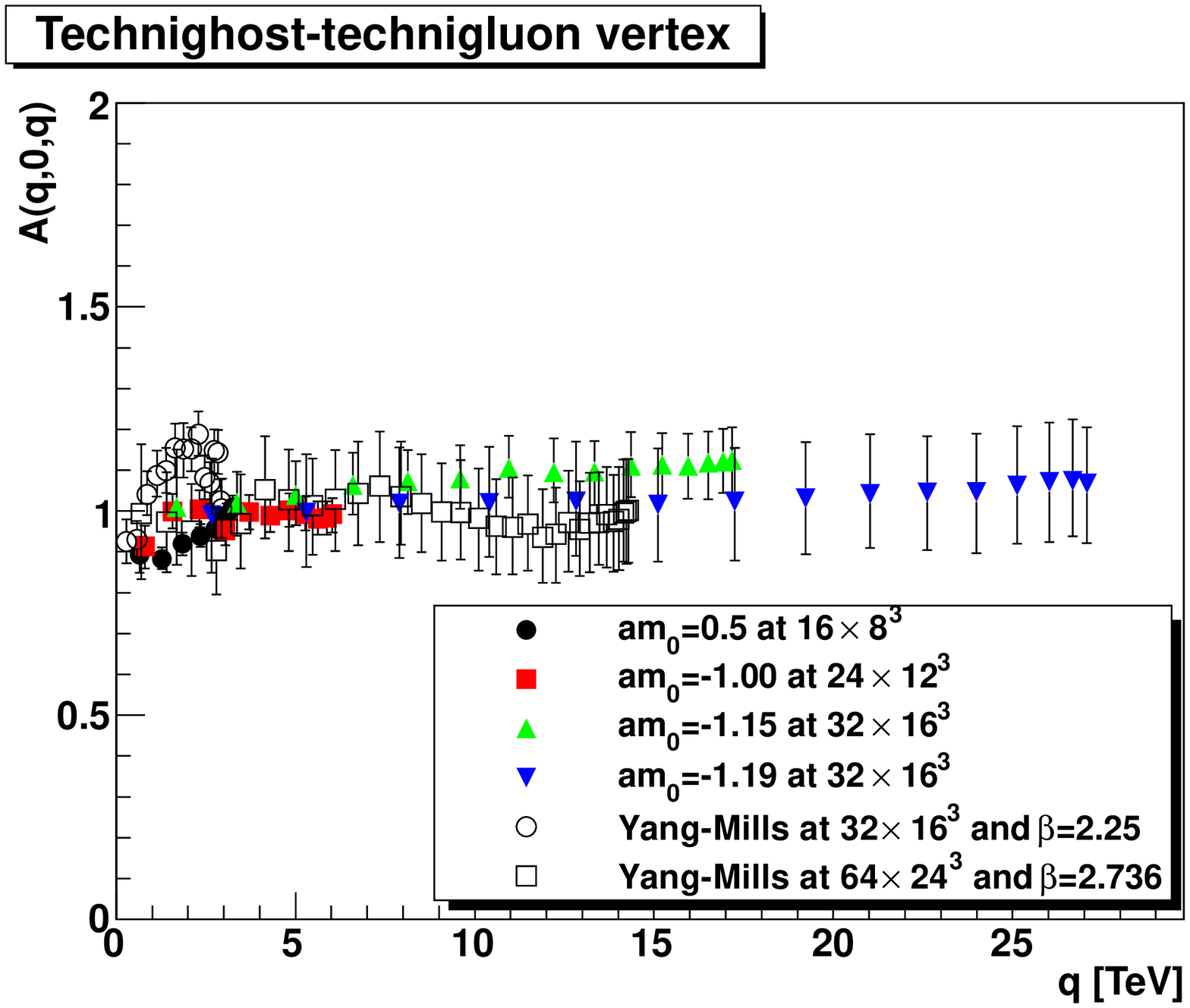}\includegraphics[width=0.5\linewidth]{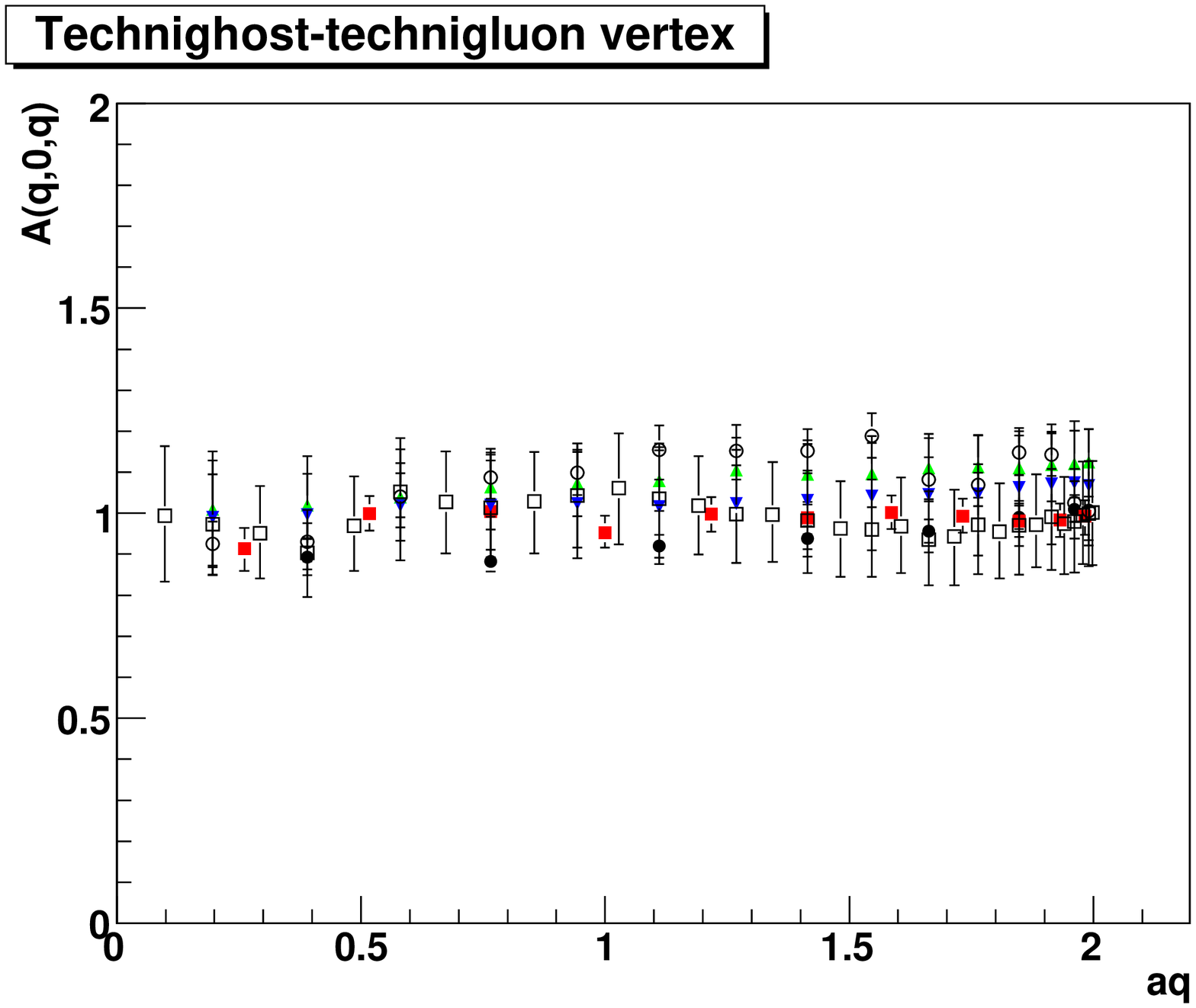}\\
\includegraphics[width=0.5\linewidth]{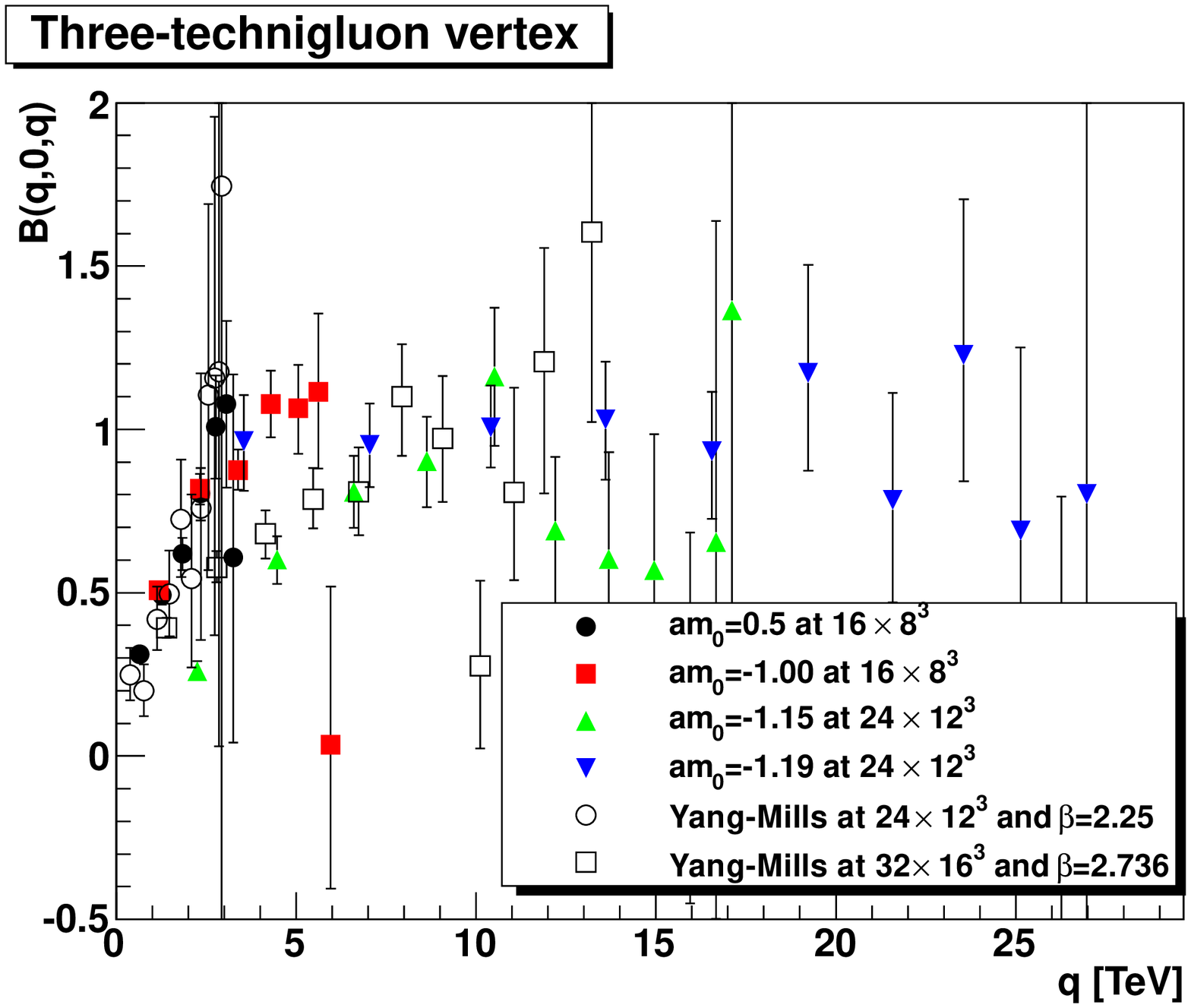}\includegraphics[width=0.5\linewidth]{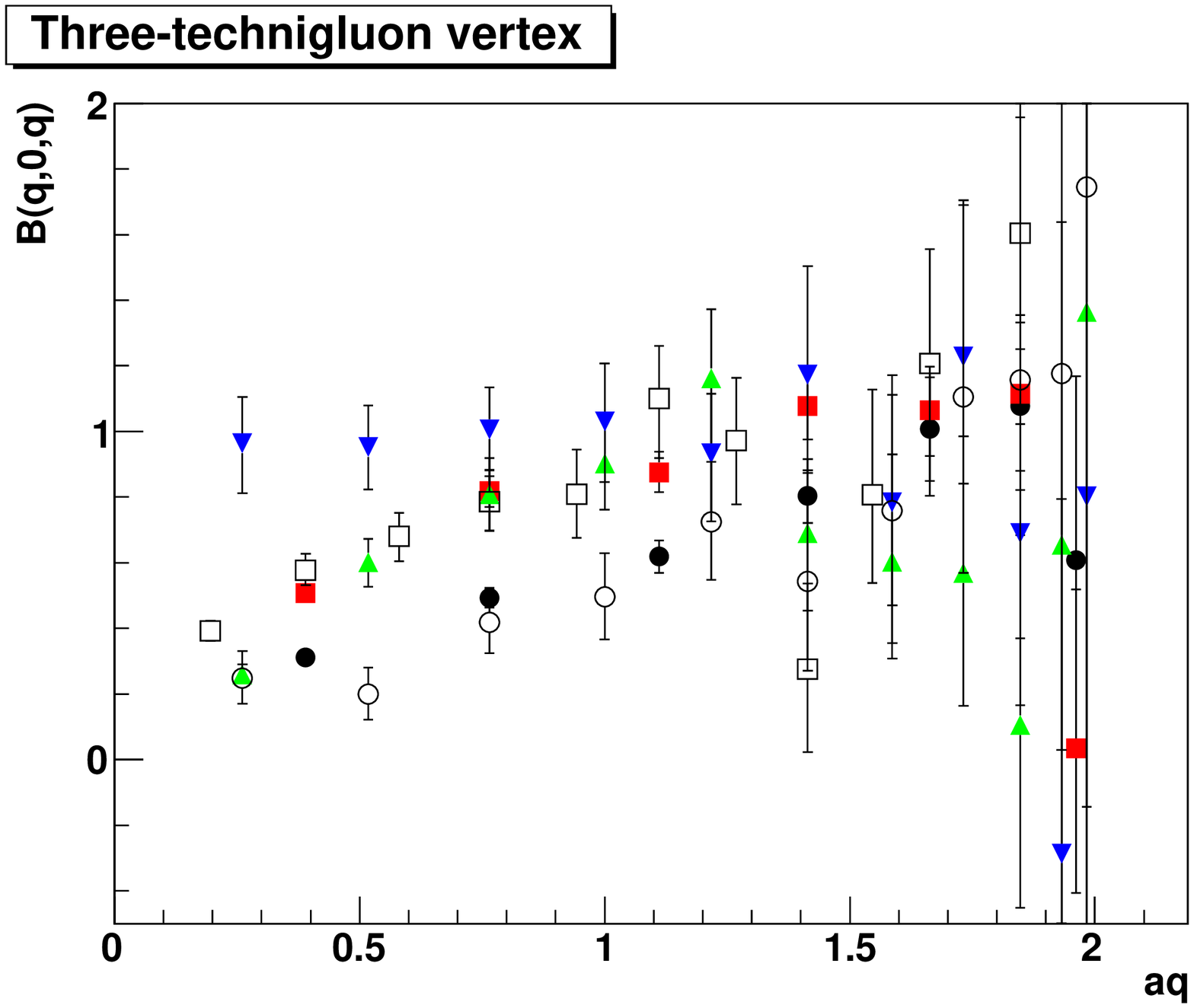}
\caption{\label{fig:vertex}The technighost-technigluon vertex with (top-left panel) and without (top-right panel) scale for vanishing technigluon momentum. The lower panels show the same for the three-technigluon vertex, with either of the momenta vanishing. Renormalization is performed at 5 TeV.}
\end{figure}

Since the running coupling is related to the three-point vertices, their behavior is interesting as well, and shown in figure \ref{fig:vertex}. The technighost-technigluon vertex is essentially momentum-independent, given the systematic uncertainties discussed in section \ref{sec:asy}. Thus, it shows no pronounced dependence on either the chosen scale nor on the techniquark mass.

The results for the three-technigluon vertex are rather difficult to assess due to the large statistical noise. Still, the trend is that the results may fall on a, more or less, universal curve for the TeV scale, while there are rather strong modifications observed for the lattice scale. In the latter case, the vertex becomes more tree-level-like with increasing techniquark mass. However, the same tendency is also observed for the Yang-Mills case when increasing $\beta$.

\section{Discussion and interpretation}\label{sec:inter}

The interpretation of the results in comparison to Yang-Mills theory is rather difficult. Lacking a solid experimental input for the scale, it is necessary to discuss two possibilities. These are the possibility that the present theory is similar to QCD, and thus that the scale is at best mildly influenced by the value of $\kappa$ \cite{Gattringer:2010zz}, or that the scale set in section \ref{sec:scale} is meaningful. Of course, this does not exhaust the possibilities, and many other options can be imagined. Here, already these two possibilities give very different interpretations, and thus the following will be restricted to these two.

The first possibility would be that the scale is essentially, up to minor quantitative corrections, controlled by $\beta$, and essentially independent of $\kappa$. Then, the results show correlation functions which become more tree-level-like with decreasing techniquark mass, and at the same time vanishing bound state masses. In particular, the interaction vertices and the running coupling become essentially momentum-independent, while the propagators behave more like the ones of photons. This would be the behavior expected for a conformal theory, where the correlation functions are essentially given by power-laws \cite{Frishman:2010zz}, which could have anomalous dimensions in four dimensions. The results here would even indicate that these anomalous dimensions are close to zero. Furthermore, the momentum-independence of the coupling constant would imply a vanishing $\beta$ function, again implying a conformal behavior. In fact, at least for the momentum range accessible here, this conformal behavior would pertain for all momenta, and the theory would not only be quasi-conformal, but genuinely conformal. This conformality would break upon coupling to the standard model \cite{Sannino:2009za}, but it is yet unclear, whether this is sufficient to make this theory useful for technicolor phenomenology.

The second possibility is essentially based on the observation that when using the scale constructed in \ref{sec:scale} all results fall on a universal curve. In particular, the changes show a behavior very similar to Yang-Mills theory when changing the scale, and the correlation functions are otherwise very similar to the Yang-Mills case. A similar behavior is not observed for changing the mass of two degenerate flavors of fundamental quarks \cite{Gattringer:2010zz}, but may be related to the presence of an infrared fixed point \cite{DeGrand:2009mt}. The simplest explanation would be the existence of a second-order phase transition when the techniquark mass vanishes. At that point, the results shown in figure \ref{fig:scale} indicate the presence of a non-zero string tension and massive techniglueballs as bound states, but massless technipions and technirhos. Since this would be at fixed gauge coupling, this would imply the existence of a non-trivial fixed point in the gauge coupling, i.\ e., the gauge coupling is not running. Since it appears rather unlikely that in the present case already a possibly existing fixed point is hit, this implies that either this is correct for any value of the gauge coupling or the present value of the gauge coupling is very close to such a critical value. This vicinity may actually be rather large if the theory would have an infrared fixed point \cite{DeGrand:2009mt}. Furthermore, the fact that the techniglueballs still have a finite mass while the correlation functions exhibit a behavior similar to Yang-Mills theories may indicate also a quasi-conformal behavior with a weakly-coupled fixed point \cite{Miransky:1998dh,Braun:2010qs}.

Note that such an interpretation is also supported by the gauge-invariant observation of the breaking of the spatial center symmetry when moving $\kappa$ closer to the critical value at fixed lattice size \cite{DelDebbio:2010hx}, as this is what would be expected for increasing $\beta$ in a Yang-Mills system at fixed lattice size. Also, the results on the number of Gribov copies shown in figure \ref{fig:gc} are in agreement with this interpretation.

At the current time, it is not possible to decide which of these interpretations (if any) is correct, due to possible lattice artifacts. Both interpretations imply a certain kind of conformal behavior, which would be in line with other observations \cite{Sannino:2009za,DelDebbio:2010hu,DelDebbio:2010hx,Bursa:2009we,DelDebbio:2009fd,Catterall:2008qk,Catterall:2007yx,Hietanen:2008mr,Hietanen:2009az,DeGrand:2009mt,DeGrand:2011qd,Lucini:2009an,Catterall:2009sb}. On the other hand, in either case the present lattice volumes and accessible discretizations and techniquark masses are not yet close to infinite volume or continuum or chiral limit for a (quasi-)conformal theory. It would be necessary to track the behavior of the theory over some orders of magnitude of momenta to be sure, and to decide whether either of the present interpretations could be correct, or if both are wrong. In particular, when it comes to taking the appropriate limits in the correct order to avoid a spurious conformal behavior \cite{Catterall:2008qk,Lucini:2009an}, a much wider range of different simulation parameters may be required.

\section{Conclusions}\label{sec:conc}

Summarizing, the (minimal Landau gauge) technigluonic two-point and three-point correlation functions have been determined for the technicolor candidate theory consisting of Yang-Mills theory coupled to two degenerate flavors of adjoint techniquarks for various techniquark masses. Within the available lattice parameters, the results show a behavior not encountered for fundamental fermions \cite{Kamleh:2007ud}. In particular, with decreasing the techniquark mass most quantities investigated, including the running gauge coupling, show quite significant effects. In lattice units, the results show a tendency to become more tree-level-like. In units of the techniglueball mass the results fall on a universal curve with decreasing lattice spacing for decreasing techniquark mass.

When comparing the results to the Yang-Mills case, it is observed that in either scale system the change of the techniquark mass can be mimicked by changing $\beta$ towards the continuum limit. If this is taken into account, the technicolor results fall within errors on a curve similar to the Yang-Mills case when changing the techniquark mass in case of the techniglueball scale. This suggests the possibility that the theory shows a second-order phase transition for vanishing techniquark mass at fixed gauge coupling, which would indicate a (quasi-)conformal behavior. Similarly, as a function of the lattice scale the behavior changes towards a more compatible one with a flat running gauge coupling, and thus also a quasi-conformal behavior. Thus, the simplest conclusion from the results here, in agreement with other results in the literature \cite{Sannino:2009za,DelDebbio:2010hu,DelDebbio:2010hx,Bursa:2009we,DelDebbio:2009fd,Catterall:2008qk,Catterall:2007yx,Hietanen:2008mr,Hietanen:2009az,DeGrand:2009mt,DeGrand:2011qd,Lucini:2009an,Catterall:2009sb}, is that the present theory could indeed be (quasi-)conformal in the limit of vanishing techniquark mass.

However, the quite significant lattice artifacts, as discussed in detail, make a final decision rather complicated. Improved results on the correlation functions would be desirable and necessary for a more solid conclusion.\\

\no{\bf Acknowledgments}\\

I am grateful to the authors of the papers \cite{DelDebbio:2010hu,DelDebbio:2010hx,DelDebbio:2009fd,DelDebbio:2008zf} for providing me with the configurations. I am furthermore grateful to Luigi Del Debbio, Christian B.\ Lang, Agostino Patella, Claudio Pica, Antonio Rago, and Francesco Sannino for helpful discussions, and to Jens Braun and Giuseppe Burgio for helpful comments on the manuscript. This work was supported by the FWF under grant number M1099-N16 and by the DFG under grant number MA 3935/5-1. Computing time was provided by the HPC cluster of the University of Graz. The ROOT framework \cite{Brun:1997pa} has been used in this project.

\bibliographystyle{bibstyle}
\bibliography{bib}

\begin{thebibliography}{10}

\bibitem{Morrissey:2009tf}
D.~E. Morrissey, T.~Plehn, and T.~M.~P. Tait,
\newblock (2009), 0912.3259.

\bibitem{Callaway:1988ya}
D.~J.~E. Callaway,
\newblock Phys. Rept. {\bf 167}, 241 (1988).

\bibitem{Bohm:2001yx}
M.~Bohm, A.~Denner, and H.~Joos,
\newblock {\em {Gauge theories of the strong and electroweak interaction}}
  (Teubner, Stuttgart, 2001),
\newblock Stuttgart, Germany: Teubner (2001) 784 p.

\bibitem{Sannino:2009za}
F.~Sannino,
\newblock Acta Phys. Polon. {\bf B40}, 3533 (2009), 0911.0931.

\bibitem{Hill:2002ap}
C.~T. Hill and E.~H. Simmons,
\newblock Phys. Rept. {\bf 381}, 235 (2003), hep-ph/0203079.

\bibitem{Lane:2002wv}
K.~Lane,
\newblock (2002), hep-ph/0202255.

\bibitem{DelDebbio:2010hu}
L.~Del~Debbio, B.~Lucini, A.~Patella, C.~Pica, and A.~Rago,
\newblock Phys.Rev. {\bf D82}, 014509 (2010), 1004.3197.

\bibitem{DelDebbio:2010hx}
L.~Del~Debbio, B.~Lucini, A.~Patella, C.~Pica, and A.~Rago,
\newblock Phys. Rev. {\bf D82}, 014510 (2010), 1004.3206.

\bibitem{Bursa:2009we}
F.~Bursa, L.~Del~Debbio, L.~Keegan, C.~Pica, and T.~Pickup,
\newblock Phys.Rev. {\bf D81}, 014505 (2010), 0910.4535.

\bibitem{DelDebbio:2009fd}
L.~Del~Debbio, B.~Lucini, A.~Patella, C.~Pica, and A.~Rago,
\newblock Phys.Rev. {\bf D80}, 074507 (2009), 0907.3896.

\bibitem{Catterall:2008qk}
S.~Catterall, J.~Giedt, F.~Sannino, and J.~Schneible,
\newblock JHEP {\bf 0811}, 009 (2008), 0807.0792.

\bibitem{Catterall:2007yx}
S.~Catterall and F.~Sannino,
\newblock Phys.Rev. {\bf D76}, 034504 (2007), 0705.1664.

\bibitem{Hietanen:2008mr}
A.~J. Hietanen, J.~Rantaharju, K.~Rummukainen, and K.~Tuominen,
\newblock JHEP {\bf 0905}, 025 (2009), 0812.1467.

\bibitem{Hietanen:2009az}
A.~J. Hietanen, K.~Rummukainen, and K.~Tuominen,
\newblock Phys.Rev. {\bf D80}, 094504 (2009), 0904.0864.

\bibitem{DeGrand:2009mt}
T.~DeGrand and A.~Hasenfratz,
\newblock Phys.Rev. {\bf D80}, 034506 (2009), 0906.1976.

\bibitem{DeGrand:2011qd}
T.~DeGrand, Y.~Shamir, and B.~Svetitsky,
\newblock (2011), 1102.2843.

\bibitem{Lucini:2009an}
B.~Lucini,
\newblock (2009), 0911.0020.

\bibitem{Catterall:2009sb}
S.~Catterall, J.~Giedt, F.~Sannino, and J.~Schneible,
\newblock (2009), 0910.4387.

\bibitem{Frishman:2010zz}
Y.~Frishman and J.~Sonnenschein,
\newblock {\em {Non-perturbative field theory: From two-dimensional conformal
  field theory to QCD in four dimensions}} (Cambridge University Press, 2010).

\bibitem{Alkofer:2000wg}
R.~Alkofer and L.~von Smekal,
\newblock Phys. Rept. {\bf 353}, 281 (2001), hep-ph/0007355.

\bibitem{Fischer:2006ub}
C.~S. Fischer,
\newblock J. Phys. {\bf G32}, R253 (2006), hep-ph/0605173.

\bibitem{Maas:2010gi}
A.~Maas,
\newblock (2010), 1011.5409.

\bibitem{Alkofer:2008et}
R.~Alkofer, C.~S. Fischer, and R.~Williams,
\newblock Eur. Phys. J. {\bf A38}, 53 (2008), 0804.3478.

\bibitem{Braun:2009gm}
J.~Braun, L.~M. Haas, F.~Marhauser, and J.~M. Pawlowski,
\newblock Phys. Rev. Lett. {\bf 106}, 022002 (2011), 0908.0008.

\bibitem{Fischer:2010fx}
C.~S. Fischer, A.~Maas, and J.~A. M{\"u}ller,
\newblock Eur. Phys. J. {\bf C68}, 165 (2010), 1003.1960.

\bibitem{Blank:2010pa}
M.~Blank, A.~Krassnigg, and A.~Maas,
\newblock Phys. Rev. {\bf D83}, 034020 (2011), 1007.3901.

\bibitem{Pawlowski:2010ht}
J.~M. Pawlowski,
\newblock (2010), 1012.5075.

\bibitem{Binosi:2009qm}
D.~Binosi and J.~Papavassiliou,
\newblock Phys. Rept. {\bf 479}, 1 (2009), 0909.2536.

\bibitem{Dudal:2010cd}
D.~Dudal, M.~S. Guimaraes, and S.~P. Sorella,
\newblock Phys. Rev. Lett. {\bf 106}, 062003 (2011), 1010.3638.

\bibitem{Elias:1984zh}
V.~Elias and M.~Scadron,
\newblock Phys.Rev.Lett. {\bf 53}, 1129 (1984).

\bibitem{Gies:2005as}
H.~Gies and J.~J\"ackel,
\newblock Eur.Phys.J. {\bf C46}, 433 (2006), hep-ph/0507171.

\bibitem{Braun:2009ns}
J.~Braun and H.~Gies,
\newblock JHEP {\bf 05}, 060 (2010), 0912.4168.

\bibitem{Aguilar:2010ad}
A.~C. Aguilar and J.~Papavassiliou,
\newblock Phys. Rev. {\bf D83}, 014013 (2011), 1010.5815.

\bibitem{Doff:2009na}
A.~Doff and A.~Natale,
\newblock Phys.Rev. {\bf D81}, 095014 (2010), 0912.1003.

\bibitem{Braun:2010qs}
J.~Braun, C.~S. Fischer, and H.~Gies,
\newblock (2010), 1012.4279.

\bibitem{DelDebbio:2008zf}
L.~Del~Debbio, A.~Patella, and C.~Pica,
\newblock Phys. Rev. {\bf D81}, 094503 (2010), 0805.2058.

\bibitem{Maas:2008ri}
A.~Maas,
\newblock Phys. Rev. {\bf D79}, 014505 (2009), 0808.3047.

\bibitem{Cucchieri:2006tf}
A.~Cucchieri, A.~Maas, and T.~Mendes,
\newblock Phys. Rev. {\bf D74}, 014503 (2006), hep-lat/0605011.

\bibitem{Fischer:2008uz}
C.~S. Fischer, A.~Maas, and J.~M. Pawlowski,
\newblock Annals Phys. {\bf 324}, 2408 (2009), 0810.1987.

\bibitem{Gribov:1977wm}
V.~N. Gribov,
\newblock Nucl. Phys. {\bf B139}, 1 (1978).

\bibitem{Singer:1978dk}
I.~M. Singer,
\newblock Commun. Math. Phys. {\bf 60}, 7 (1978).

\bibitem{Cucchieri:1995pn}
A.~Cucchieri and T.~Mendes,
\newblock Nucl. Phys. {\bf B471}, 263 (1996), hep-lat/9511020.

\bibitem{Maas:2010qw}
A.~Maas,
\newblock JHEP {\bf 02}, 076 (2011), 1012.4284.

\bibitem{vonSmekal:2009ae}
L.~von Smekal, K.~Maltman, and A.~Sternbeck,
\newblock Phys. Lett. {\bf B681}, 336 (2009), 0903.1696.

\bibitem{vonSmekal:1997vx}
L.~von Smekal, A.~Hauck, and R.~Alkofer,
\newblock Ann. Phys. {\bf 267}, 1 (1998), hep-ph/9707327.

\bibitem{Cvitanovic:2008}
P.~Cvitanovi\'c,
\newblock {\em {Group theory}} (Princeton University Press, Princeton, 2008),
\newblock Princeton University Press (2008).

\bibitem{Cucchieri:2004sq}
A.~Cucchieri, T.~Mendes, and A.~Mihara,
\newblock JHEP {\bf 12}, 012 (2004), hep-lat/0408034.

\bibitem{Ball:1980ax}
J.~S. Ball and T.-W. Chiu,
\newblock Phys. Rev. {\bf D22}, 2550 (1980).

\bibitem{Maas:2007uv}
A.~Maas,
\newblock Phys. Rev. {\bf D75}, 116004 (2007), 0704.0722.

\bibitem{Cucchieri:2006za}
A.~Cucchieri and T.~Mendes,
\newblock Phys. Rev. {\bf D73}, 071502 (2006), hep-lat/0602012.

\bibitem{Cucchieri:2007ta}
A.~Cucchieri, A.~Maas, and T.~Mendes,
\newblock Phys. Rev. {\bf D75}, 076003 (2007), hep-lat/0702022.

\bibitem{Silva:2005hb}
P.~J. Silva and O.~Oliveira,
\newblock Phys. Rev. {\bf D74}, 034513 (2006), hep-lat/0511043.

\bibitem{Cucchieri:2008fc}
A.~Cucchieri and T.~Mendes,
\newblock Phys. Rev. {\bf D78}, 094503 (2008), 0804.2371.

\bibitem{Cucchieri:2007rg}
A.~Cucchieri and T.~Mendes,
\newblock Phys. Rev. Lett. {\bf 100}, 241601 (2008), 0712.3517.

\bibitem{Bornyakov:2009ug}
V.~G. Bornyakov, V.~K. Mitrjushkin, and M.~M{\"u}ller-Preussker,
\newblock Phys. Rev. {\bf D81}, 054503 (2010), 0912.4475.

\bibitem{Bogolubsky:2009dc}
I.~L. Bogolubsky, E.~M. Ilgenfritz, M.~M{\"u}ller-Preussker, and A.~Sternbeck,
\newblock Phys. Lett. {\bf B676}, 69 (2009), 0901.0736.

\bibitem{Fischer:2007pf}
C.~S. Fischer, A.~Maas, J.~M. Pawlowski, and L.~von Smekal,
\newblock Annals Phys. {\bf 322}, 2916 (2007), hep-ph/0701050.

\bibitem{Cucchieri:2008qm}
A.~Cucchieri, A.~Maas, and T.~Mendes,
\newblock Phys. Rev. {\bf D77}, 094510 (2008), 0803.1798.

\bibitem{Schleifenbaum:2004id}
W.~Schleifenbaum, A.~Maas, J.~Wambach, and R.~Alkofer,
\newblock Phys. Rev. {\bf D72}, 014017 (2005), hep-ph/0411052.

\bibitem{Ilgenfritz:2006he}
E.~M. Ilgenfritz, M.~M{\"u}ller-Preussker, A.~Sternbeck, A.~Schiller, and I.~L.
  Bogolubsky,
\newblock Braz. J. Phys. {\bf 37}, 193 (2007), hep-lat/0609043.

\bibitem{Fischer:2009tn}
C.~S. Fischer and J.~M. Pawlowski,
\newblock Phys. Rev. {\bf D80}, 025023 (2009), 0903.2193.

\bibitem{Alkofer:2008jy}
R.~Alkofer, M.~Q. Huber, and K.~Schwenzer,
\newblock Phys. Rev. {\bf D81}, 105010 (2010), 0801.2762.

\bibitem{Cucchieri:1997ns}
A.~Cucchieri,
\newblock Nucl. Phys. {\bf B521}, 365 (1998), hep-lat/9711024.

\bibitem{Cucchieri:2010ft}
A.~Cucchieri and T.~Mendes,
\newblock PoS {\bf LATTICE2010}, 280 (2010), 1101.4537.

\bibitem{Kamleh:2007ud}
W.~Kamleh, P.~O. Bowman, D.~B. Leinweber, A.~G. Williams, and J.~Zhang,
\newblock Phys. Rev. {\bf D76}, 094501 (2007), 0705.4129.

\bibitem{Maas:2009se}
A.~Maas,
\newblock Phys. Lett. {\bf B689}, 107 (2010), 0907.5185.

\bibitem{Zwanziger:1993dh}
D.~Zwanziger,
\newblock Nucl. Phys. {\bf B412}, 657 (1994).

\bibitem{Bornyakov:2008yx}
V.~G. Bornyakov, V.~K. Mitrjushkin, and M.~M\"uller-Preussker,
\newblock Phys. Rev. {\bf D79}, 074504 (2009), 0812.2761.

\bibitem{Maas:2010wb}
A.~Maas,
\newblock PoS {\bf LATTICE2010}, 279 (2010), 1010.5718.

\bibitem{Cucchieri:1997dx}
A.~Cucchieri,
\newblock Nucl. Phys. {\bf B508}, 353 (1997), hep-lat/9705005.

\bibitem{Maas:2010nc}
A.~Maas,
\newblock Eur. Phys. J. {\bf C71}, 1548 (2011), 1007.0729.

\bibitem{Sternbeck:2008mv}
A.~Sternbeck and L.~von Smekal,
\newblock Eur.Phys.J. {\bf C68}, 487 (2010), 0811.4300.

\bibitem{Gattringer:2010zz}
C.~Gattringer and C.~B. Lang,
\newblock {\em Quantum chromodynamics on the lattice} (Lect. Notes Phys.,
  2010).

\bibitem{Miransky:1998dh}
V.~Miransky,
\newblock Phys.Rev. {\bf D59}, 105003 (1999), hep-ph/9812350.

\bibitem{Brun:1997pa}
R.~Brun and F.~Rademakers,
\newblock Nucl. Instrum. Meth. {\bf A389}, 81 (1997).

\end{thebibliography}


\end{document}